\documentclass[aps,preprint,showpacs,preprintnumbers]{revtex4}

\usepackage{graphicx}
\usepackage{epsfig}
\usepackage{amsmath}
\usepackage{times}
\usepackage{bm}
\def\@biblabel#1{{#1.}} %    %  LV \def\@BIBLABEL#1{$^{#1}\m@th$} %
\def\be{\begin{equation}} \def\ee{\end{equation}}
\newcommand{\ket}[1]{\mbox{$|#1\rangle$}}
\newcommand{\bra}[1]{\mbox{$\langle#1|$}}
\begin{document}
\normalsize

% \begin{document}

\title{Electromagnetically induced transparency in superconducting
quantum circuits : Effects of decoherence, tunneling and multi-level
cross-talk}

\author{Zachary Dutton$^1$, K.~V.~R.~M. Murali$^2$, William D. Oliver$^3$, and T.~P. Orlando$^2$}
\affiliation{${}^1$Naval Research Laboratory, Washington, DC 20375 \\
${}^2$Department of Electrical Engineering and Computer Science \\
Massachusetts Institute of Technology, Cambridge MA 02138 \\
${}^3$MIT Lincoln Laboratory, 44 Wood Street, Lexington, MA 02420}

\begin{abstract}
We explore theoretically electromagnetically-induced transparency
(EIT) in a superconducting quantum circuit (SQC). The system is a
persistent-current flux qubit biased in a $\Lambda$ configuration.
Previously [Phys. Rev. Lett. {\bf 93}, 087003 (2004)], we showed
that an ideally-prepared EIT system provides a sensitive means to
probe decoherence. Here, we extend this work by exploring the
effects of imperfect dark-state preparation and specific decoherence
mechanisms (population loss via tunneling, pure dephasing, and
incoherent population exchange). We find an initial, rapid
population loss from the $\Lambda$ system for an imperfectly
prepared dark state. This is followed by a slower population loss
due to both the detuning of the microwave fields from the EIT
resonance and the existing decoherence mechanisms. We find analytic
expressions for the slow loss rate, with coefficients that depend on
the particular decoherence mechanisms, thereby providing a means to
probe, identify, and quantify various sources of decoherence with
EIT. We go beyond the rotating wave approximation to consider how
strong microwave fields can induce additional off-resonant
transitions in the SQC, and we show how these effects can be
mitigated by compensation of the resulting AC Stark shifts.

\end{abstract}

\date{\today}

\maketitle

\section{\label{sec:intro} Introduction}

Superconducting quantum circuits (SQCs) based on Josephson
junctions (JJs) exhibit macroscopic quantum-coherent
phenomena~\cite{Leggett85a}. These circuits exhibit quantized flux
or charge states states, depending on their fabrication
parameters. The quantized states are analogous to the quantized
internal (hyperfine and Zeeman) levels in an atom, and the SQCs
thus behave like ``artificial atoms.''
Spectroscopy~\cite{Friedman00a,Wal00a,Berkley03a,Xu05a}, Rabi
oscillations and Ramsey
interferometry~\cite{Nakamura99a,Nakamura01a,Vion02a,Yu02a,Martinis02a,Chiorescu03a,Claudon04a},
cavity quantum electrodynamics~\cite{Chiorescu04a,Wallraff04a},
and St\"{u}ckelberg oscillations~\cite{Oliver05a,Sillanpaa05a} are
examples of quantum-mechanical behavior first realized in atomic
systems that have also been recently demonstrated with SQCs.

We recently leveraged the atom-SQC analogy to propose
electromagnetically induced transparency
(EIT)~\cite{Boller91a,Harris97a} in superconducting
circuits~\cite{EIT}. EIT has attracted much attention in atomic
systems in the context of slow light~\cite{Nature1},
%WDO,OtherUSL}
quantum
memory~\cite{stoppedLight,stoppedLightTheory,quantumStorage} and
nonlinear optics~\cite{NLO}. EIT occurs in so-called
``$\Lambda$-systems'' comprising two meta-stable states, each
coupled via resonant electromagnetic fields to a third, excited
state. For particular initial states called ``dark states,'' the
absorption on {\it both} transitions is suppressed due to
destructive quantum interference, thus making the atom transparent
to the applied fields.   Though EIT is often studied in the
context of the behavior of a weak `probe' field in the presence of
a stronger 'pump' field, we focus on the case where the two fields
have comparable amplitude. In Ref.~\cite{EIT}, we analyzed a
superconducting persistent-current qubit biased such that it
exhibited a $\Lambda$-configuration: two meta-stable states (the
qubit) and a third, shorter-lived state (the readout state). We
showed that EIT provides a non-destructive means to confirm
preparation of an arbitrary superposition state of the qubit.
Moreover, we showed that the proposed EIT scheme can sensitively
probe the qubit decoherence rate using a method analogous to the
proposal in Ref.~\cite{Ruostekoski99a} for atomic systems.  This
method compliments other available techniques of probing
decoherence such as spin echo \cite{spinEcho} and Rabi oscillation
decay \cite{microRes}. Because the EIT method requires no
manipulation of the qubit during the probing, it offers unique
advantages in this regard. In addition to our EIT work, several
groups have considered the use of ``dark states'' in SQCs
comprising a $\Lambda$-configuration to implement adiabatic
passage and its application to quantum information
processing~\cite{Han_list,Amin03a,Paspalakis_list}.

In the present work, we extend and augment our analysis in
Ref.~\cite{EIT} with realistic effects which arise in SQCs due to
the presence of additional quantized levels (beyond the three-level
``$\Lambda$-system'' model). These effects have qualitatively unique
signatures in an EIT experiment, and this work provides a tool for
identifying their origin. This allows a more complete understanding
of the full level-structure of the SQC system, and it further
clarifies the necessary criteria for the experimental observation of
EIT. The present work carries the spirit of previous investigations
in which additional degrees of freedom (beyond two-level models)
were required to explain quantitatively experimental Rabi
oscillations in SQCs. Examples of these works include resonant
tunneling across the barrier~\cite{resCoup}, diagonal dipole matrix
elements~\cite{diagonal}, and coupling to additional degrees of
freedom outside the SQC, such as micro-resonators
\cite{microRes,microResThoery}. Just as EIT is sensitive to
decoherence, it will be similarly sensitive to effects beyond the
idealized three-level model.

The effects we investigate arise primarily from differences between
SQCs and the atomic systems considered in much of the literature.
First, while damping of the excited level is provided naturally by
spontaneous emission in atoms, in SQCs, this decay is `manufactured'
by resonant biasing across the tunnel barrier followed by fast
measurement with a SQUID. This process must be considered in more
detail to assure this decay is indeed analogous to spontaneous decay
in atoms. Second, the transitions are at microwave rather than
optical frequencies, whereas the Rabi frequency coupling rates and
dephasing rates tend to be faster than in atomic systems.  This
means that various couplings in the system can be more comparable to
the level spacings and thus the rotating wave approximation (RWA) is
often not as valid as in atomic systems. Third, the level structure
itself is quite different. In particular, there is typically some
degree of dipole-like coupling between {\it all} pairs of levels in
the system, because selection rules allow all possible
transitions~\cite{selectionRules}. Fourth, in SQCs, there is the
possibility of direct resonant tunneling across the barrier, a
feature which is absent in atomic systems.

This paper is organized in the following manner. In
Section~\ref{sec:pcqubit}, we introduce the proposed system, a
persistent-current (PC) qubit \cite{Mooij99a,Orlando99a}. We discuss
the conditions under which the PC qubit exhibits a
$\Lambda$-configuration amongst its multi-level energy band
structure that is conducive for an EIT demonstration.  We then
present the Hamiltonian and density matrix approaches to analyze the
system dynamics. In Section~\ref{sec:EIT}, we use the Hamiltonian
approach to give useful analytic approximations to the full system,
which allow us to investigate EIT in a reduced three-level system.
We explore effects of population and phase mismatch between the
prepared initial state and the desired dark state (as defined by the
applied fields), and the effect of detuning the applied fields from
their resonances. We also consider the SQUID measurement rate and
its effect on the effective decay and frequency of the excited
`read-out' level. In Section~\ref{sec:loss}, we use the density
matrix approach to include pure dephasing and incoherent population
loss and exchange, generalizing our previous results in
Ref.~\cite{EIT}. We explore the effect on EIT in the presence of
coherent and incoherent tunneling processes. Generally, one must
make the EIT `preparation rate' (proportional to the microwave field
intensities) faster than the decoherence rate in order to observe
EIT. In Section~\ref{sec:crosstalk}, we go beyong the rotating-wave
approximation (RWA) to examine the important issue of microwave
field-induced off-resonant transitions in the spirit of previous
work on two-level systems~\cite{RWA}. We conclude that off-resonant
transitions cause frequency shifts and losses which depend on the
coupling field intensities. Unlike decoherence and detuning, these
transitions generally manifest themselves as the field intensities
are increased. The high-field frequency shifts are analogous to the
AC-stark shifts observed in atomic systems. We show how the
off-resonant effects can be mitigated and, in some cases, eliminated
by compensating for the frequency shifts.

\section{The PC-qubit}
\label{sec:pcqubit}

\subsection{The PC-qubit model}
\label{subsec:SQCstructure}

\begin{figure}
\includegraphics[width=3in]{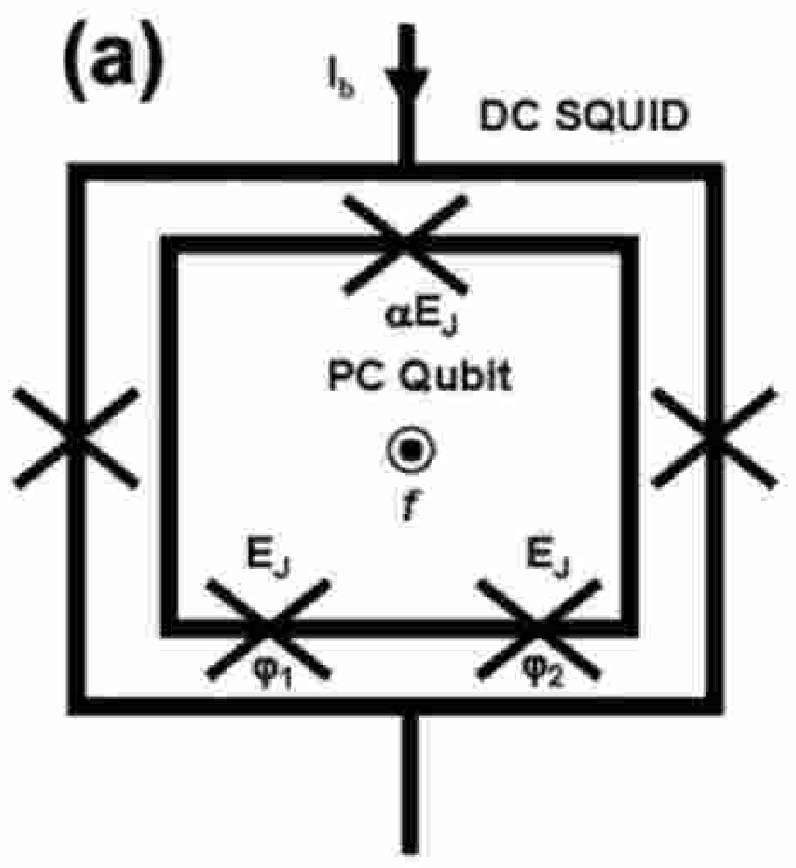}
\includegraphics[width=3in]{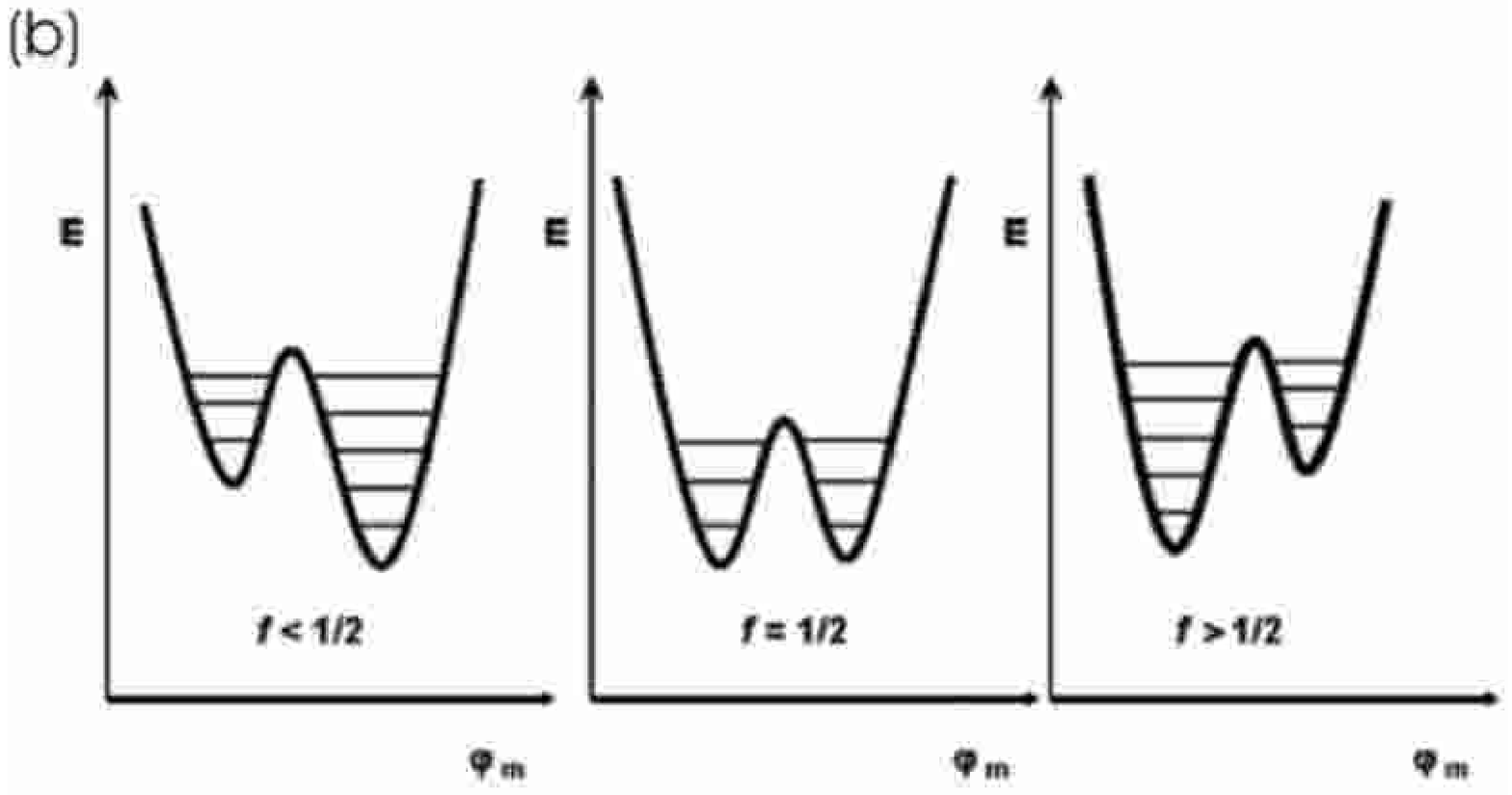}
\includegraphics[width=3.5in]{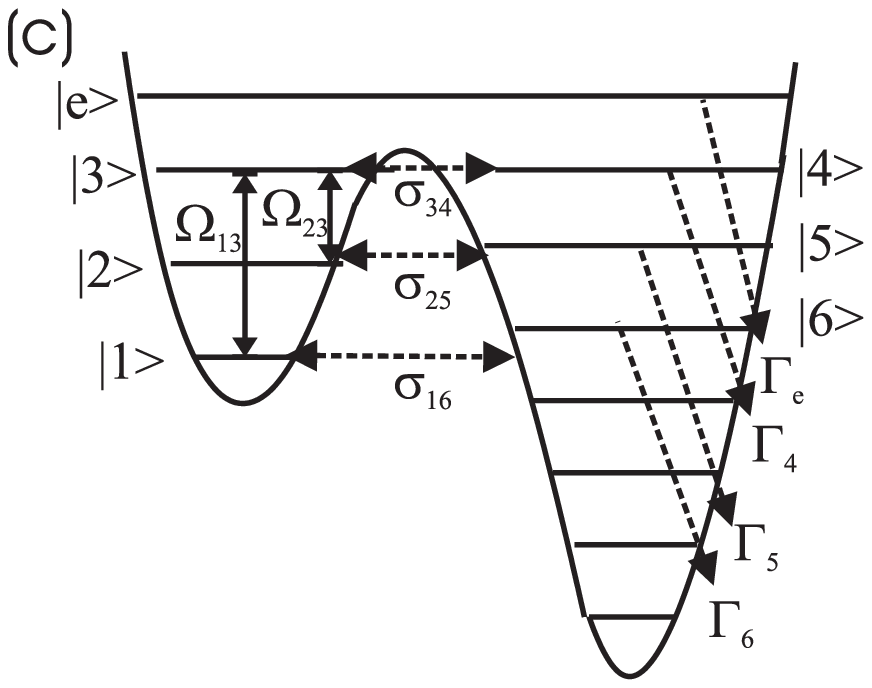}
\caption{\label{fig:pcqubit}\textbf{A pc-qubit with a dc SQUID
measuring device} \textbf{(a)} A pc-qubit, a superconducting loop
with two Josephson junctions of equal dimension and the third scaled
by a factor $\alpha$, as shown in the inner loop. The outer loop is
a dc-SQUID that is used to measure the magnetic moment of the qubit.
\textbf{(b)} A representation of the potential energy of the
pc-qubit as a function of $f$, the magnetic flux in the loop in
units of the flux quantum $\Phi_0$. The qubit potential can range
from an asymmetric double well biased to the right to a symmetric
double well and to the an symmetric double well biased to the left
for $f$ ranging from $< 1/2$, $=1/2$, and $> 1/2$ respectively.
\textbf{(c)} One-dimensional double-well potential and energy-level
diagram for $f=0.502$, in which case we have a three-level system in
the left-hand well.  States $\ket{1}$ and $\ket{2}$ are meta-stable,
while $\ket{3}$ will have significant loss via resonant tunneling to
$\ket{4}$ ($\sigma_{34}$).  The right-hand well states undergo fast
damping ($\Gamma_4,\Gamma_5,\Gamma_6$) via the SQUID measurement and
intrawell relaxation to lower states. Coupling between the our
three-levels is induced by two resonant microwave fields with Rabi
frequencies $\Omega_{13}$ and $\Omega_{23}$, forming the
$\Lambda$-system.
%The calculated pc-qubit
%parameters~\cite{Orlando99a} are,
The qubit parameters we use in calculations are
$\omega_2-\omega_1=(2 \pi)~27.8~\mathrm{GHz}$ and
$\omega_3-\omega_2=(2 \pi)~27~\mathrm{GHz}$, with matrix elements
$x_{ij}$ for $(i,j)=$ (1,2), (2,3), and (1,3) set to -0.0145,
-0.0371, -0.0263, respectively.}
\end{figure}

The persistent-current (PC) qubit is a superconductive loop
interrupted by two Josephson junctions of equal size and a third
junction scaled smaller in area by the factor $0.5 < \alpha < 1$
(Fig.~\ref{fig:pcqubit}(a))~\cite{Orlando99a,Mooij99a}. Its
dynamics are described by the Hamiltonian
\begin{multline}
 \label{eq:pcq}
 {\cal{H}}_{pc} = \frac{1}{2}C
 \left( \frac{\Phi_{0}}{2\pi} \right)^{2}
 (\dot{\varphi}^{2}_{p} + (1 + 2 \alpha) \dot{\varphi}^{2}_{m}) \\
  + E_{j} \left[2 + \alpha - 2 \cos\varphi_{p} \cos\varphi_{m} -
 \alpha \cos(2 \pi f + 2 \varphi_{m})\right],
\end{multline}
in which $C$ is the capacitance of the larger junctions,
$\varphi_{p,m} \equiv (\varphi_{1} \pm \varphi_{2})/2$,
$\varphi_{i}$ is the gauge-invariant phase across the larger
junctions $i=\{ 1,2 \}$, $E_j= I_c \Phi_0/2 \pi$ is the Josephson
coupling energy, $I_c$ is the critical current of the larger
junctions, and $f$ is the magnetic flux through the loop in units
of the flux quantum $\Phi_{0}$~\cite{Orlando99a}.

The qubit potential energy (the second term in ${\cal{H}}_{pc}$)
forms forms a 2D periodic double well potential, a one-dimensional
slice through which is shown in Fig.~\ref{fig:pcqubit}(b). Each well
corresponds to a distinct classical state of the electric current,
{\textit{i.e.,}} left or right circulation through the loop, with a
net magnetization of that is discernable using a dc
SQUID~\cite{Orlando99a}.  The relative depth of the two wells can be
adjusted by detuning the flux bias to either side of the symmetric
point $f=1/2$.  The potential wells exhibit quantized energy levels
corresponding to the quantum states of the macroscopic circulating
current~\cite{Segall03a,Crankshaw03a,Yu04a,Yu05a}, with the number
of levels on each side determined by the depth and frequencies of
the wells. In this basis the Hamiltonian can be written

\begin{align}
 \label{eq:Hsys}
  {\cal{H}}_{pc} &=\mathcal{H}_0 + \mathcal{H}_\mathrm{tunnel}; \\
 \label{eq:H0} \mathcal{H}_0 &= \hbar \sum_i \omega_i \ket{i}\bra{i}, \\
 \label{eq:Htunnel}
 \mathcal{H}_\mathrm{tunnel}&= \hbar \sum_{i,j\not=i} \sigma_{ij} \ket{i}\bra{j}.
\end{align}
We note here that two points of view may be taken when discussing
the system described by the Hamiltonian in Eq.~(\ref{eq:Hsys}). In
one picture, the diabatic states (diagonal matrix elements) of the
qubit are the uncoupled single-well states of classical circulating
current, and these diabatic states are coupled through the tunneling
terms (off-diagonal matrix elements). In a second picture, the
Hamiltonian is diagonalized, resulting in eigenenergies and
eigenstates of the double-well potential. Although the perspectives
differ, these two pictures will, of course, lead to identical
results; only the interpretation differs. Throughout the paper, we
primarily describe the dynamics in terms of diabatic (single-well)
states coupled through the tunneling barrier of the double-well
potential and driven by harmonic excitation. Exceptions, where they
exist, will be clearly noted.

The three-level $\Lambda$ structure to implement EIT is then
provided by the left-hand meta-stable states $\ket{1},\ket{2}$ and
the fast decaying level $\ket{3}$ shown in
Fig.~\ref{fig:pcqubit}(c). Each of these levels are taken to have a
finite loss rate $\Gamma_i^{\mathrm{(t)}}$, due to resonant
tunneling to a right-well state (at $\sigma_{ij}$) followed by
relaxation of the right-hand well state $\Gamma_j$ (which is a sum
of population relaxation to lower levels and damping induced by a
fast SQUID measurement of the circulation current of right-hand well
states). In particular, we desire a fast decay rate
$\Gamma_3^{\mathrm{(t)}}$, which is achieved by resonantly biasing
$\ket{3}$ and $\ket{4}$ and a fast SQUID measurement ($\approx $
1-10~ns), as analyzed in Section~\ref{subsec:measurement}.
Conversely, we desire states $\ket{1}$ and $\ket{2}$ to be
long-lived and the tunneling $\sigma_{25}$ will cause loss and
decoherence, which is analyzed in Section~\ref{subsec:tunelling}
($\sigma_{16}$ is negligible by comparison). Rough estimates of the
interwell loss rates for {\it resonant}-tunneling are
$\Gamma_1^{\mathrm{(t)}} \approx (1~\mathrm{ ms})^{-1}, \,
\Gamma_2^{\mathrm{(t)}} (1 \; \mu\mathrm{s})^{-1}$, and
$\Gamma_3^{\mathrm{(t)}} \approx (1~\mathrm{ns})^{-1}$ and the
off-resonant biasing of states $\ket{1} -\ket{6}$ and
$\ket{2}-\ket{5}$ will significantly decrease these rates. In
addition, states $\ket{2}$ and $\ket{3}$ can have intrawell
relaxation rates $\Gamma_{3\rightarrow 1}, \, \Gamma_{3\rightarrow
2}, \, \Gamma_{2\rightarrow 1}$ (not shown in the diagram). Under
similar bias conditions the rate $\Gamma_{3\rightarrow 1} +
\Gamma_{3\rightarrow 2} \approx (25 \; \mu\text{s})^{-1}$
(experimentally measured in ~\cite{Yu04a}) is much slower than
$\Gamma_3^{(t)}$.  Also note that, $\Gamma_{2 \rightarrow 1}$,
another source of decoherence of the meta-stable states, will be
less than $\Gamma_{3\rightarrow 1} + \Gamma_{3\rightarrow 2}$.

These quantized levels may be coupled using microwave radiation.  An
applied radiation field $\mu$ can be described in terms of an
amplitude, frequency and phase: $\Delta f_\mu = g_\mu
\cos(\omega_\mu t + \phi_\mu)$. We find the resulting matrix
elements for level transitions (the Rabi frequencies) by treating
$\Delta f_\mu$ as a small perturbation in the $\cos(2 \pi f + 2
\varphi_{m})$ term in Eq.~(\ref{eq:pcq}).  We write it as $\sin(2
\pi f + 2 \varphi_{m}) \sin(2 \pi \Delta f_\mu)$, which can be
approximated as $\sin(2 \pi f + 2 \varphi_{m}) (2 \pi \Delta
f_\mu)$, leading to a Rabi frequecy $\Omega_{ij}^{(\mu)} \equiv (2
\pi) g_\mu \alpha E_{j} x_{ij}/\hbar$, where $x_{ij}\equiv
\bra{i}\sin(2 \pi f + 2 \varphi_{m}) \ket{j}$.  The elements
$x_{ij}$ we calculate for our proposed parameters are listed in the
caption of Fig.~\ref{fig:pcqubit}(c). In EIT, we address the SQC
with two microwave fields $\Delta f_a = g_a \cos(\omega_a t+
\phi_a)$, with $\omega_a \approx \omega_3-\omega_1$, and $\Delta f_b
= g_b \cos(\omega_b t+\phi_b)$, with $\omega_b \approx
\omega_3-\omega_2$. The microwave induced Hamiltonian can then be
written as:
\begin{equation}
\label{eq:Hmw} \mathcal{H}_{\mu-\mathrm{wave}}= \frac{\hbar}{2}
\sum_{i,j} \sum_\mu \big(\Omega_{ij}^{(\mu)}  e^{-i (\omega_\mu
t+\phi_\mu)}+c.c. \big) \ket{i}\bra{j}
\end{equation}
where $i,j$ runs over the states and $\mu$ runs over the two fields
$a,b$. We emphasize that the above approximation is a perturbative
approach valid only for small driving amplitudes. In the strongly
driven limit, the approximation breaks down, preventing the Rabi
frequency from growing without bound~\cite{Nakamura01a,Oliver05a}.

Microwave excitation is used to establish the population of
meta-stable states (such as $\ket{1}$ and $\ket{2}$) via {\it
photon-assisted tunneling}. In this scheme, the population of a
meta-stable state is driven via a resonant radiation field into a
{\it read-out state} ({\it i.e.} $\ket{3}$) which quickly tunnels
to a {\it measurement state} ({\it i.e.} $\ket{4}$). This state
has opposing current circulation with a unique flux signature that
can be measured using a DC-SQUID~\cite{Orlando99a}. In the present
scheme, {\it two} fields are simultaneously applied (resonant with
$\ket{1} \leftrightarrow \ket{3}$ and $\ket{2} \leftrightarrow
\ket{3}$; see Fig.~\ref{fig:pcqubit}(c)), and EIT is manifested by
a suppression of the photon-assisted tunneling due to quantum
interference between the two excitation processes.

\subsection{Evolution model}
\label{subsec:model}

It is convenient to calculate dynamics from the above Hamiltonian
terms in an interaction picture which transforms away the diagonal
energies $\mathcal{H}_0$ (\ref{eq:H0}).  In this frame the total
Hamiltonian is then the sum of (\ref{eq:Htunnel}) and
(\ref{eq:Hmw}):

\begin{eqnarray}
\label{eq:Hint} \tilde{\mathcal{H}} = \frac{\hbar}{2}
\sum_{i,j}\sum_\mu \big( \Omega_{ij}^{(\mu)} \, e^{-i (\omega_\mu
t+\phi_\mu)}+c.c.\big) e^{i (\omega_i-\omega_j)t}\ket{i}\bra{j} +
\hbar \sum_{i,j\not=i} \sigma_{ij} e^{i
(\omega_i-\omega_j)t}\ket{i}\bra{j}
\end{eqnarray}

\noindent  Note that the exponential arguments involve sums of
microwave frequencies $\omega_\mu$ and level splittings
$\omega_i-\omega_j$.  When these nearly cancel the state is said
to be near-resonant and the coupling is strong. However, for most
of the terms, this cancellation does not occur, and the term
rotates its phase rapidly on the scale of frequencies of interest
($\Omega_{ij}^{(l)}$,  $\sigma_{ij}$ and $\Gamma_j$). Such terms
are neglected in the Rotating Wave Approximation (RWA).

We can also include incoherent losses from the levels, $\Gamma_i$,
by introducing an additional non-Hermitian part of the Hamiltonian
$\tilde{\mathcal{H}}_{\mathrm{relax}}=- i \hbar \sum_i
(\Gamma_i/2)\ket{i}\bra{i}$. This is often done in quantum
optics~\cite{Scully,thesis} to include non-Hermitian decay of
radiatively decaying levels. We then describe the system by a
wavefunction $\ket{\tilde{\Psi}}=\sum_{i}
\tilde{c}_{i}(t)\ket{i}$, with the initial population normalized
to unity $\sum_{i} |\tilde{c}_{i}(0)|^{2}=1$ (this can decay in
time due to the non-Hermitian loss).   The evolution of the
$\ket{\tilde{\Psi}}$ is the governed from Schr\"odinger's
equation:
\begin{eqnarray}
\label{eq:sch} i \hbar \frac{\partial}{\partial
t}\ket{\tilde{\Psi}}=(\tilde{\mathcal{H}}+\tilde{\mathcal{H}}_\mathrm{relax})\ket{\tilde{\Psi}}.
\end{eqnarray}

\noindent  Besides giving the coherent dynamics, this
Schr\"odinger equation correctly predicts the population
relaxation of level $\ket{i}$ at $\Gamma_i$ and also gives the
correct dephasings of coherences between $\ket{i}$ and other
states at half this rate $\Gamma_i/2$.

When necessary, we use a  density matrix approach to include
incoherent processes.  For example, pure dephasing of a coherence
between the two meta-stable states $\ket{1}$ and $\ket{2}$ goes
beyond the Hamiltonian approach (\ref{eq:sch}). Similarly,
incoherent feeding of levels (such as population into $\ket{1}$
from interwell relaxation $\Gamma_{2 \rightarrow 1}$) goes beyond
this description.  The density matrix is written
$\tilde{\rho}=\sum_{ij} \tilde{\rho}_{ij} \ket{i} \bra{j}$, where
the population in the levels are given by the diagonal terms
$\tilde{\rho}_{ii}$ and correspond to $|\tilde{c}_i|^2$ in the
wavefunction description, while the off-diagonal terms
$\tilde{\rho}_{ij}$ correspond to $\tilde{c}_i \tilde{c}_j^*$ and
describe coherences between levels.  The evolution of the density
matrix is given by

\begin{eqnarray}
\label{eq:dmEvol} i \hbar \frac{\partial}{\partial
t}\tilde{\rho}=[\tilde{\mathcal{H}}+\tilde{\mathcal{H}}_\mathrm{relax},\tilde{\rho}]
+ \mathcal{L} \tilde{\rho}.
\end{eqnarray}

\noindent The first term reproduces the part already predicted by
the Schr\"odinger equation (\ref{eq:sch}), while the
super-operator $\mathcal{L}$, the Lindbladian \cite{Scully},
accounts for other incoherent processes. For pure dephasing of the
$\ket{i}\leftrightarrow \ket{j}$ coherence, $\gamma_{ij}$, we
introduce a term $\mathcal{L}_{ij,ij}=-\gamma_{12}$.  For a
population relaxation from $\ket{j} \rightarrow \ket{i}$,
 $\Gamma_{j \rightarrow i}$, we introduce $\mathcal{L}_{jj,ii}=+\Gamma_{j\rightarrow
 i}$.  The associated population {\it loss} from $\ket{j}$ and decoherences
are already included through Eq.~\ref{eq:sch}  (via a term $-i \hbar
(\Gamma_{j \rightarrow i}/2)\ket{i}\bra{i}$).

Throughout the paper, we consider the model in a number of distinct
cases.  In each, we include three levels $\ket{1},\ket{2},\ket{3}$,
coupled by two microwave fields $\Delta f_a , \, \Delta f_b$ making
up our $\Lambda$ system (see Fig.~\ref{fig:pcqubit}(c)).  We then
selectively include additional levels, such as the $\ket{4}, \,
\ket{5},$ and $\ket{e}$ to isolate the contributions of each of
them. Numerical results were obtained with a fourth-order
Runge-Kutta algorithm \cite{numRec} solving Eq.~(\ref{eq:dmEvol}).
In it we do not make any RWA assumptions {\it a priori}, but instead
introduce some cut-off frequency $\omega_{RWA}$.  We examine the
phase factors of each term in the evolution and set to zero ones
with phases rotating faster than $\omega_{RWA}$.

We compare our numerical results with approximate analytic
solutions in many cases.  When possible, we use the Schr\"odinger
equation (\ref{eq:sch}) to obtain simpler analytic results, though
the full density matrix approach is used when dephasing and
interwell relaxation are considered (Sections~\ref{subsec:bloch}-
\ref{subsec:incoherent}). In the analytic results we normally make
an additional transformation $\ket{\tilde{\Psi}} \rightarrow
\ket{\Psi}, \,
\tilde{\mathcal{H}}+\tilde{\mathcal{H}}_\mathrm{relax} \rightarrow
\mathcal{H}$, defined by  transformations of each level frequency
$\{\tilde{c}_i\} \rightarrow \{c_i\}
 = \{\tilde{c}_i e^{i \delta_i t}\}$,
where the $\delta_i$ are chosen in to eliminate time-dependent
exponential phase factors in (\ref{eq:Hint}) (they are usually {\it
detunings}, that is, frequency mismatches between the microwave
frequencies and the corresponding transitions). Detuning from
two-photon resonance, decoherence, and additional levels are all
seen to destroy the perfect transparency of EIT and cause slow
exponential loss of the population. We will obtain expressions for
the loss rate $R_L$ in these cases.

\section{Electromagnetically Induced Transparency in a SQC}
\label{sec:EIT}

\subsection{Ideal EIT in a $\Lambda$ configuration}
\label{subsec:ideal}

We first consider the `ideal' case in which the three levels in
the left well (see Fig.~\ref{fig:pcqubit}(c)) are well isolated
from direct tunneling to other levels, states $\ket{1}$ and
$\ket{2}$ are perfectly stable, and $\ket{3}$ quickly decays at
some fast rate $\Gamma_3^{(\mathrm{t})}$. This decay is in reality
due to resonant coupling of $\ket{4}$ ($\sigma_{34}$) and
subsequent SQUID measurement $\Gamma_4$, but we will see in
Section~\ref{subsec:measurement} how one can derive
$\Gamma_3^{(\mathrm{t})}$ in terms of these underlying processes.

We apply fields with nearly resonant frequencies
$\omega_a=\omega_3-\omega_1+ \Delta_{13}$ and
$\omega_b=\omega_3-\omega_2+\Delta_{23}$ (see
Fig.~\ref{fig:pcqubit}(c)), where the $\Delta_{13},\Delta_{23}$ are
small detunings. All other couplings are sufficiently detuned to
safely eliminate them under the RWA.  In this case the
transformations to eliminate phase rotating terms are given by
$\delta_1=0, \, \delta_2=\Delta_{13}-\Delta_{23}, \,
\delta_3=\Delta_{13}$.   The Hamiltonian, written in matrix notation
in a basis $\{\ket{1},\ket{2},\ket{3}\}$ is

\begin{eqnarray}
\label{eq:Hideal} \mathcal{H} = \frac{\hbar}{2}
\left[\begin{matrix}
0 & 0 & \Omega_{13}^{*} \\
0 & - 2 \Delta_{2} & \Omega_{23}^{*} \\
\Omega_{13} & \Omega_{23} & -i \Gamma_3^{(\mathrm{t})} - 2
\Delta_{13}
\end{matrix} \right]
\end{eqnarray}

\noindent where $\Delta_2 \equiv \Delta_{13}-\Delta_{23}$ is the
detuning from {\it two-photon resonance}.  Here we have dropped
the $a,b$ labels, $\Omega_{13} \equiv \Omega_{13}^{(a)}$ and
$\Omega_{23} \equiv \Omega_{23}^{(b)}$ (see Eq.~(\ref{eq:Hint}))
as there is no ambiguity.  The open system loss of $\ket{3}$ due
to tunneling $\Gamma_3^{(\mathrm{t})}$ is assumed to dominate
incoherent population exchange due to intra-well relaxation,
allowing a Schr\"odinger evolution analysis (\ref{eq:sch}).

First consider the resonant case $\Delta_{13}=\Delta_{23}=0$. A
qubit initially in the ground state $\ket{1}$ can be prepared in a
superposition state $\ket{\Psi_\mathrm{init}}=c_{1}\ket{1} +
c_{2}\ket{2}$ by temporarily driving it with a field resonant with
the $\ket{1} \leftrightarrow \ket{2}$ transition. Applying only one
field $\Omega_{13}$ ($\Omega_{23}$) field then allows the population
of a state $\ket{1}$ ($\ket{2}$) to be read out through a transition
to state $\ket{3}$ followed by a rapid escape to the right well. In
this case, the superposition is destroyed by the absorption of a
photon.

However, from (\ref{eq:sch}) and (\ref{eq:Hideal}) it follows that
if we simultaneously apply both fields and the SQC is in the {\it
dark}  state

\begin{equation}
\label{eq:darkState}
\ket{\Psi_D}=\frac{\Omega_{23}}{\Omega}\ket{1} -
\frac{\Omega_{13}}{\Omega}\ket{2},
\end{equation}

\noindent (where $\Omega \equiv
\sqrt{|\Omega_{13}|^2+|\Omega_{23}|^2}$) then (\ref{eq:sch})
predicts $\dot{c}_1=\dot{c}_2=\dot{c}_3=0$. For this particular
state, the two absorption processes, $\ket{1} \leftrightarrow
\ket{2}$ and $\ket{1} \leftrightarrow \ket{3}$, have equal and
opposite probability amplitude and thus cancel by quantum
interference.  As a result, no excitation into $\ket{3}$, and thus
no tunneling to the right well, will be observed. Note that
$\ket{\Psi_D}$ constrains both the relative intensity {\it and}
phase of the light fields. Any other (non-zero) values for the
relative amplitudes in the two states $\ket{1},\ket{2}$ will lead to
a coupling into $\ket{3}$ and subsequent loss.

An alternative interpretation is obtained by examining the
eigensystem of the Hamiltonian (\ref{eq:Hideal}).  The dark state
$\ket{\Psi_D}$ has eigenvalue zero. The other two eigenstates are
linear combinations of the excited state $\ket{3}$ and the
combination of the stable states orthogonal to $\ket{\Psi_D}$:
$\ket{\Psi_A}=(\Omega_{13}^*\ket{1}+\Omega_{23}^*\ket{2})/\Omega$,
called the {\it absorbing state} (a ``bright state'').    The
system $\{\ket{\Psi_A}, \ket{3}\}$ acts effectively as a two-level
system coupled by $\Omega$.   The eigenvalues corresponding to the
two eignestates are $(- i \Gamma_3^{(\mathrm{t})} \pm
\sqrt{4\Omega^2-\Gamma_3^{(\mathrm{t})2}})/4$ and the imaginary
parts of these eigenvalues give the loss rates of these states. In
the limit $\Omega \gg \Gamma_3^{(\mathrm{t})}$, these rates are
both $\Gamma_3^{(\mathrm{t})}/4$ and one observes damped Rabi
oscillations. In the limit $\Omega \ll \Gamma_3^{(\mathrm{t})}$,
there is an eigenstate $\approx \ket{\Psi_A}$ with a slower
damping rate $\Omega^2/2\Gamma_3^{(\mathrm{t})}$.

\begin{figure}
\includegraphics{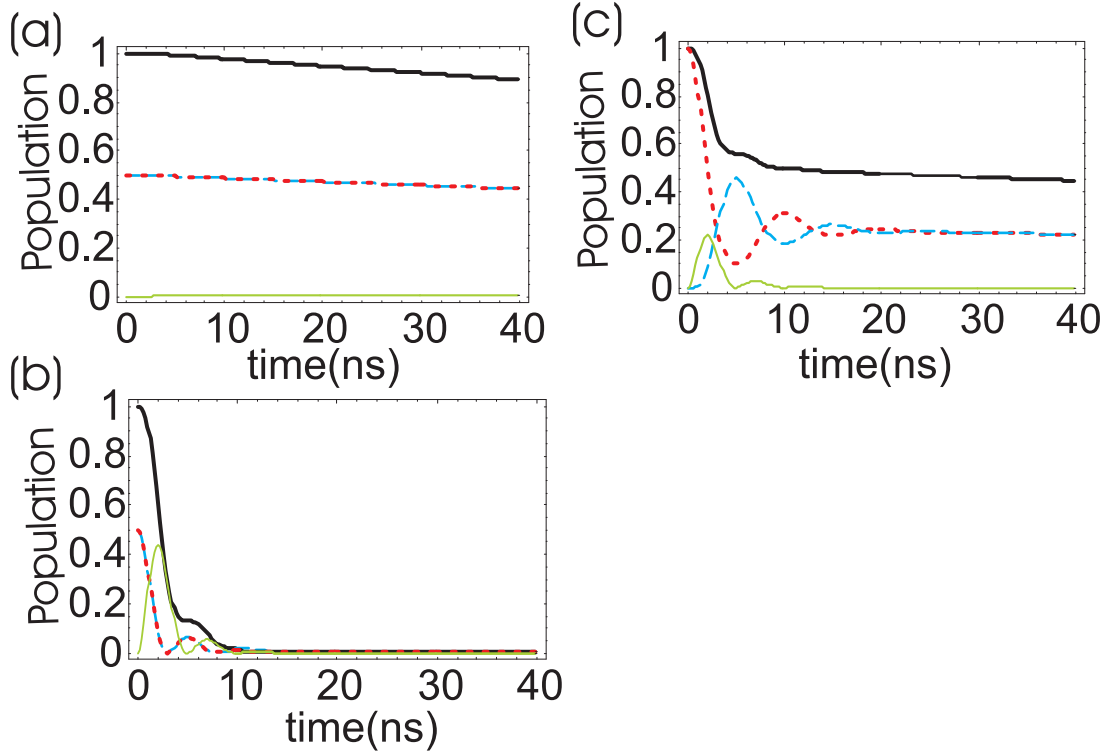}
\caption{\label{fig:outPhase}(Color online) \textbf{Suppression of
tunneling due to EIT for various ideal wavefunctions}  \textbf{(a)}
The populations of the states as a function of time in the presence
of applied fields $\Omega_{13}=\Omega_{23}=(2 \pi)~150~\mathrm{MHz}$
and tunneling rate $\Gamma_3^{(\mathrm{t})}=(2 \pi)~130~\mathrm{MHz}
= 1/1.2$~ns for the initial state $\rho_{11}=\rho_{22}=0.5 \, ,
\rho_{12}=-0.5$ (the dark state). The dotted (red) curve shows
$\rho_{11}$, the dashed (blue) $\rho_{22}$ and the thin solid
(green) $\rho_{33}$. The total population (sum of the three) is the
thick solid (black) curve. There is a slow exponential decay of the
population due to the dephasing rate $\gamma_{12}=(2 \pi)1~$MHz.
\textbf{(b)} The population evolutions (same convention) for the
initial state $\rho_{11}=\rho_{22}=0.5 \, , \rho_{12}=0.5$ (the
absorbing state). \textbf{(c)} The population evolutions for an
initial state $\ket{1}$ (which is an equal superposition of the dark
and absorbing states).}
\end{figure}

Figure~\ref{fig:outPhase}(a) shows an example of the lack of
tunneling in the presence of applied fields
$\Omega_{13}=\Omega_{23}$ for the corresponding dark state
$\ket{\Psi_{\mathrm{init}}} = (\ket{1}-\ket{2})/\sqrt{2}$ ({\it
i.e.} $\rho_{11}=\rho_{22}=0.5 \, , \rho_{12}=-0.5$). One sees only
a barely perceptible population $\rho_{33}$ and a very slow loss of
the $\rho_{11}$ and $\rho_{22}$.  This is due to a pure dephasing of
the state coherence, which we take to be $\gamma_{12}=(2
\pi)~1$~MHz.  The effect of this dephasing is a small exponential
loss at a rate we label $R_L^{(\gamma_{12})}$, which is discussed
and derived in \cite{EIT} and reviewed in
Section~\ref{subsec:dephasing}. Otherwise the populations remain
$\rho_{11}=\rho_{22}\approx 0.5$. EIT thus provides a means to
confirm, without disturbing the system, that one had indeed prepared
the qubit in a particular desired state of the SQC, preserving its
quantum coherence.

By contrast, Fig.~\ref{fig:outPhase}(b) shows the large loss induced
when one applies these same fields to the absorbing state, {\it
i.e.}, the state with the same populations but $\pi$ out of phase:
$\ket{\Psi_{\mathrm{init}}}=(\ket{1}+\ket{2})/\sqrt{2}$. In
Fig.~\ref{fig:outPhase}(b) we see that there is a large population
in the $\ket{3}$ and the entire population has tunneled to the right
well within about 10~ns.   Note that here we are in the intermediate
regime $\Omega \sim \Gamma_3^{(\mathrm{t})}$ so we get oscillations
with period $\sim \Omega$ strongly damped at $\sim
\Gamma_3^{(\mathrm{t})}/2$. This is the completely analogous to the
tunneling which occurs with a single applied field in a two-level
scheme.

A general state can be decomposed into dark and absorbing state
components. Fig.~\ref{fig:outPhase}(c) shows a case where the
initial population is purely in $\ket{1}$ and the same fields are
applied. Here the initial state can be written
$\ket{\Psi_{\mathrm{init}}}=\ket{1}=(\ket{\Psi_\mathrm{D}}+\ket{\Psi_\mathrm{A}})/\sqrt{2}$.
Half of the population (the component in the absorbing state) is
coupled out over the 10~ns time scale while the dark state
component remains. In terms of level populations $\rho_{11}, \,
\rho_{22}$, approximately 1/4 of the population is coherently
coupled from $\ket{1}$ to $\ket{2}$.

\subsection{EIT with imperfect state preparation}
\label{subsec:imperfectState}

One of the useful aspects of EIT is the extremely sensitive manner
in which it can measure the amplitude and phase of superpositions
in the SQC.  When the prepared state has a slightly different
phase or population ratio than the state we intend to prepare, EIT
could be used to measure these deviations. Such imperfect
preparation could arise, for example, due to imperfections in the
preparation pulse.

\begin{figure}
\includegraphics[width=3in]{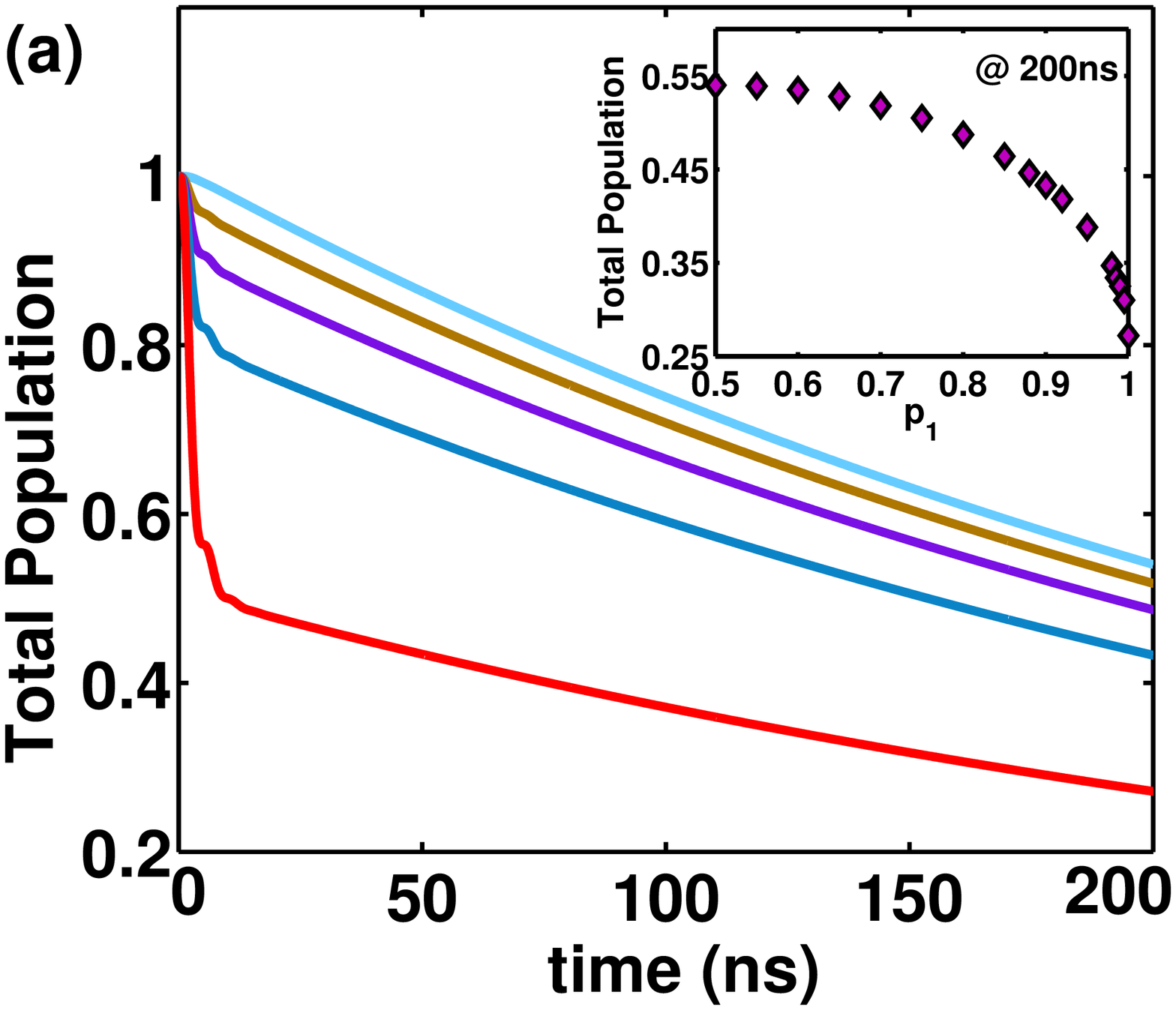}
\includegraphics[width=3in]{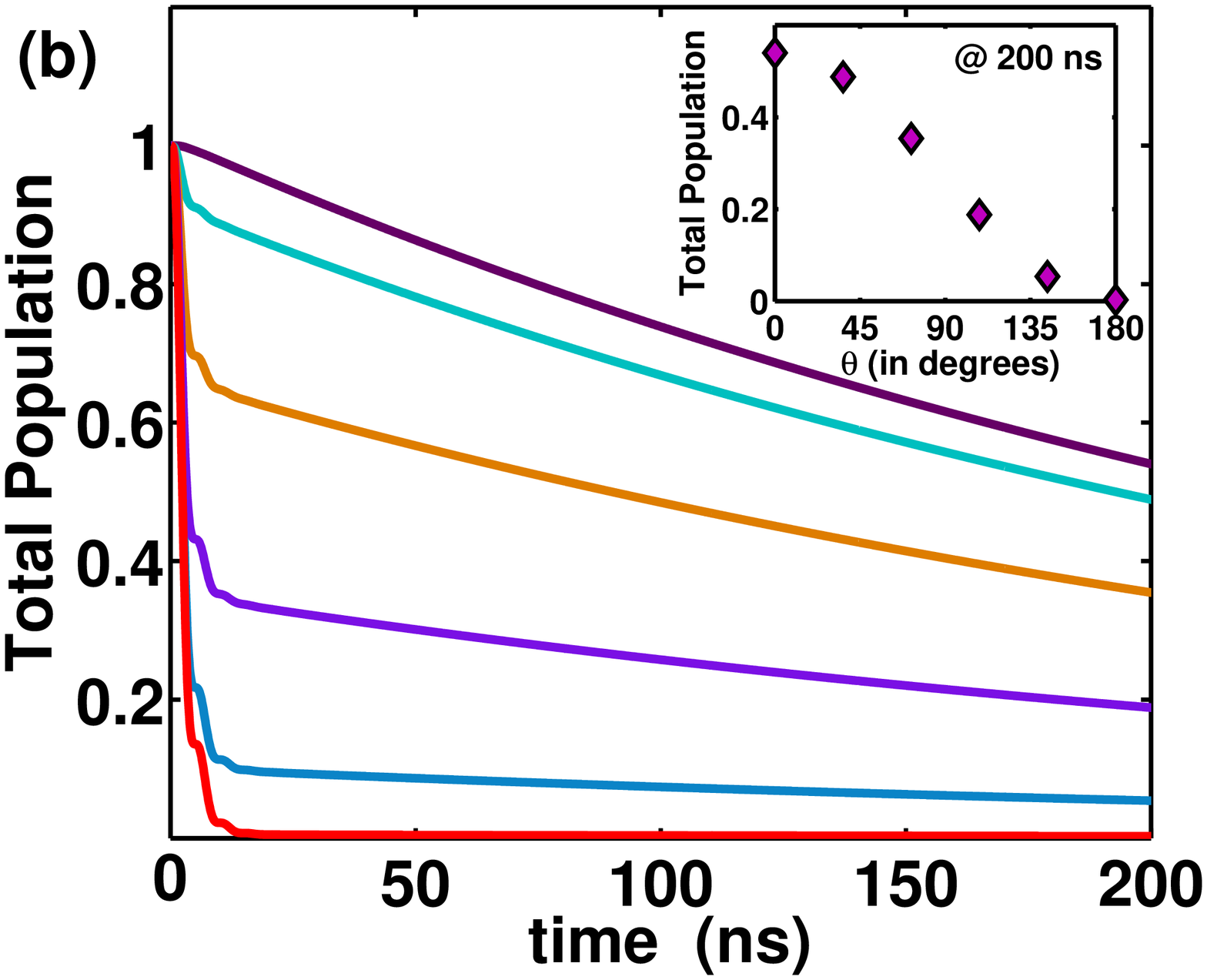}
\caption{\label{fig:imperfectState} (Color online)\textbf{ Imperfect
state preparation}\textbf{(a)} The time evolution of the population
in the left well as a function of initial state of the form
$\ket{\Psi_\mathrm{init}}= \sqrt{p_{1}} \ket{1} - \sqrt{(1-p_{1})}
\ket{2}$. The uppermost curve is for the dark state $p_{1} = 0.5$.
Successively lower curves are for $p_1=$0.6, 0.7, 0.8, 0.9 and 1.0.
There is a sharp initial decay when there are deviations from the
dark state. Inset: The population in the left well at 200~ns versus
$p_1$. \textbf{(b)} The population decay out of the left well as a
function of the initial phase of the prepared state,
$\ket{\Psi_\mathrm{init}} = (\ket{1} - e^{i \theta} \ket{2} )/
\sqrt{2}$. $\theta=0$ (top curve) and the other curves are $\pi/5,
\, 2\pi/5, \, 3\pi/5, \, 4\pi/5$, and $\pi$. We see full decay for
$\theta=\pi$, the absorbing state $\ket{\Psi_A} = (\ket{1} + \ket{2}
)/ \sqrt{2}$. Inset: The population in the left well at 200 ns as a
function of $\theta$. }
\end{figure}

Fig.\ref{fig:imperfectState}(a), shows the populations loss when
preparing a state with various initial state population ratios and
applying fields $\Omega_{13}=\Omega_{23}$.  We again introduce a
small dephasing $\gamma_{12}=(2 \pi)1$~MHz.  The inset shows the
population at 200~ns, well after the initial transient losses have
occured.   This data can be understood using the dark/absorbing
basis discussed above.   The modulus square of the overlap of the
initial state and the dark state $\langle \Psi_D
\ket{\Psi_\mathrm{init}}$ gives the population remaining after the
fast initial loss of the absorbing component.   Postulating that
the slower loss (due to dephasing or other effects) is exponential
with some rate $R_L$, the population after the fast initial loss
is:

\begin{equation}
\label{eq:pop} |\langle \Psi_\mathrm{init} \ket{\Psi_{D}}|^2
e^{-R_L t}.
\end{equation}

\noindent Thus, detecting the fast initial decay of the population
can indicate the mismatch between the fields and the prepared
state population.

Phase mismatch, or unwanted $z$-rotation, in the qubit preparation
shows similar behavior.   Figure~\ref{fig:imperfectState}(b) shows
the population decay from the left well for the state $(\ket{1} -
e^{i\theta} \ket{2})/ \sqrt{2}$. The upper most line is decay due
to the perfect state, while the lower lines indicate the decay for
varying value of $\theta$.   This example indicates that the dark
state, is more sensitive to phase mismatch than population
mismatch.

\subsection{EIT detuned from resonance}
\label{subsec:detuning}

Because EIT is a coherent effect, it only occurs in a narrow range
of frequencies near the two-photon resonance.  The width of the EIT
feature is generally determined by the field intensities, and can be
made narrower than the broad resonances of the individual one-photon
transitions ($\ket{1} \leftrightarrow \ket{3}$ and $\ket{2}
\leftrightarrow \ket{3}$), which are determined by the fast decay
rate of $\ket{3}$ $\Gamma_3^{(\mathrm{t})}$.

\begin{figure}
\includegraphics[width=3in]{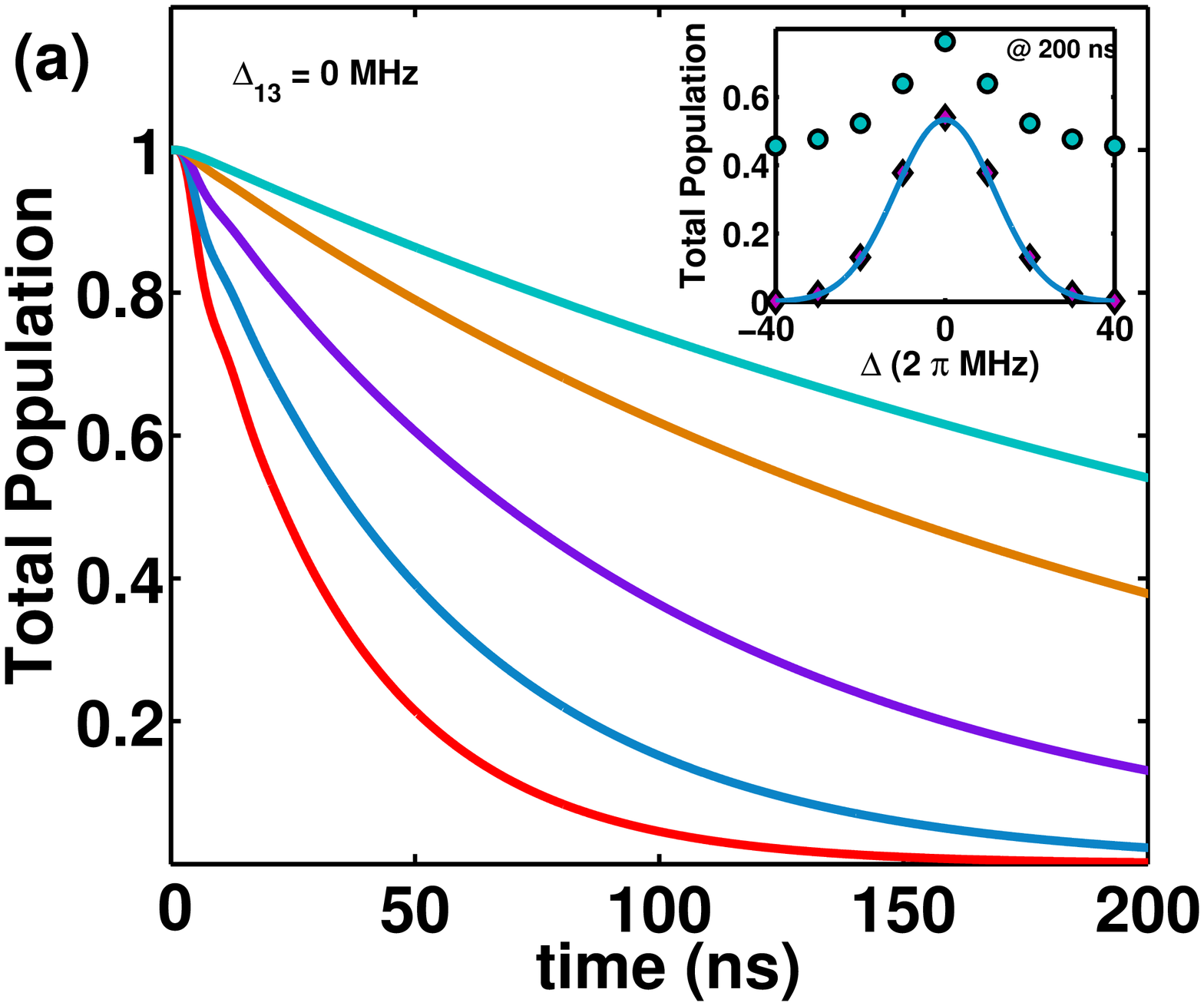}
\includegraphics[width=3in]{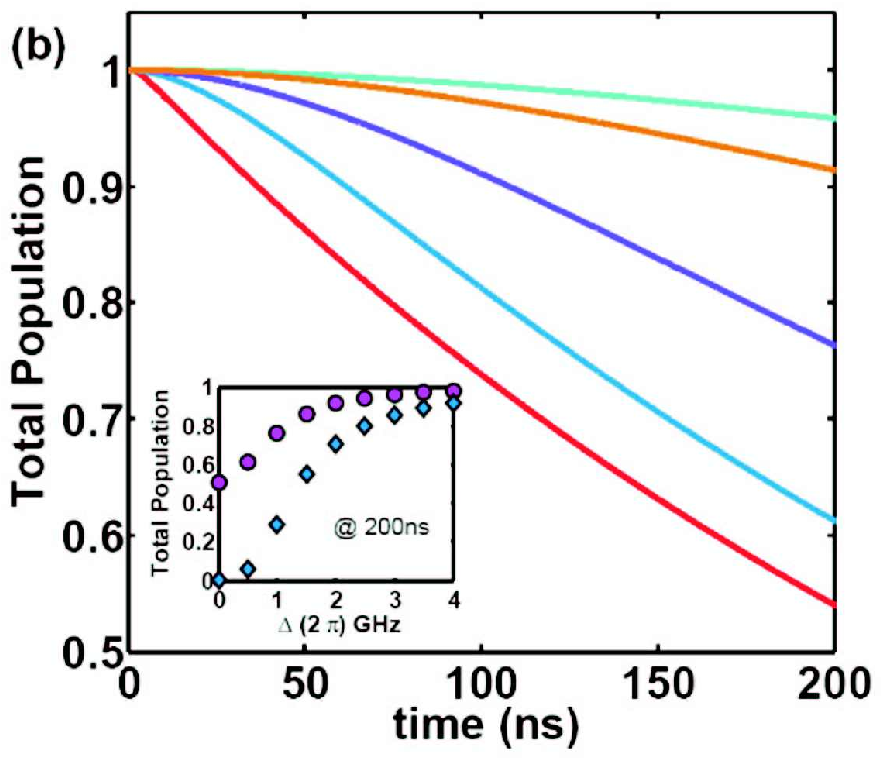}
\caption{\label{fig:withDetune} (Color online) \textbf{EIT in the
presence of detuning.} \textbf{(a)} Numerical calculation of the
total population in time when $\Delta_{13}=0$ MHz, and at various
$\Delta_{23}$ (top to bottom curve) $(2 \pi)$~0, 10, 20, 30, 40~MHz.
For $\Delta_{23}=0$, the decay is due to pure dephasing, while the
decay is sharper when $\Delta_{23} \not= 0$. Inset: The population
in the left well at 200 ns as a function of the two-photon detuning
$\Delta_{2} \equiv \Delta_{13}-\Delta_{23}$, with $\Delta_{13} = 0$
(circles). The solid curve shows the prediction (\ref{eq:detLoss}).
The diamonds show the case $\Delta_{13} = (2 \pi)1$~GHz, varying
$\Delta_{2}$ about the two-photon resonance. \textbf{(b)} The
population decay as a function one-photon detuning at two-photon
resonance, $\Delta_{13} = \Delta_{23} = \Delta $ for (top to bottom)
$(2 \pi)$~3,
 2, 1, 0.5, 0~GHz. As the one-photon detuning increases, the effective
coupling of the fields to the transition reduces, hence reducing
the rate of decay.   Inset: The population in the left well at 200
ns as a function of the detuning for the dark state (circles). For
comparison we also show the population remaining for the absorbing
state $\ket{\Psi_\mathrm{init}}=(\ket{1}+\ket{2})/\sqrt{2}$ at the
same detunings (diamonds). }
\end{figure}

Figure~\ref{fig:withDetune}(a) shows the results of simulations
with the same parameters as Fig.~\ref{fig:outPhase}(a), but with
the detuning $\Delta_{23}$ varied.  The curves show exponential
loss occuring at various rates.  We can analyze the results with
the Hamiltonian (\ref{eq:Hideal}) and the corresponding
Schr\"odinger equation (\ref{eq:sch}).  We first adiabatically
eliminate \cite{adiabatic} the excited level by setting
$\dot{c}_3=0$ and obtain

\begin{equation}
\label{eq:c3det} c_3 =
(\Omega_{13}c_1+\Omega_{23}c_2)\bigg(\frac{2 \Delta_{13} - i
\Gamma_3^{(\mathrm{t})}}{4 \Delta_{13}^2 +
\Gamma_3^{(\mathrm{t})2}}\bigg)
\end{equation}

\noindent  This expression is valid for times long compared to the
initial transient time
$\mathrm{Min}\{(\Gamma_3^{(\mathrm{t}))-1},\Delta_{13}^{-1}\}$.
Note that for the dark state (\ref{eq:darkState}) the amplitude
$c_3$ vanishes. Plugging this expression back into the equations
for $\dot{c}_1, \, \dot{c}_2$ then gives a $2 \times 2$ matrix
evolution equation, which can be easily solved by finding for its
eigenvalues and eigenvectors. For $\Delta_2=0$ the eigenvectors
are simply the dark $\ket{\Psi_D}$ and absorbing $\ket{\Psi_A}$
states of Section~\ref{subsec:ideal}, with eigenvalues $\lambda_D
= 0$ and $\lambda_A = -\Omega^2 (\Gamma_3^{(\mathrm{t})}- 2 i
\Delta_{13})/ 2(4 \Delta_{13}^2+\Gamma_3^{(\mathrm{t})2})$,
respectively.  The absorbing component population is damped out at
$-2 Re\{\lambda_A\}$.  In many cases, we are interested in the
regime close to the one-photon resonance $\Delta_{13} \ll
\Gamma_3^{(\mathrm{t})}$ for which this reduces to
$\Omega^2/\Gamma_3^{(\mathrm{t})}$ as in
Section~\ref{subsec:ideal}.

With a non-zero two-photon detuning $\Delta_2$, this process is
complicated an additional phase evolution term.  Plugging
(\ref{eq:Hideal}) into Schr\"odinger's equation (\ref{eq:sch}) (in
the frame defined before (\ref{eq:Hideal})) gives a term
$\dot{c}_2= \dots + i \Delta_2 c_2$, which acts to drive the phase
of the SQC out of the dark state and competes with damping of the
absorbing component. Solving for the eigensystem in this case we
see that, in the limit of small two-photon detuning ($|\Delta_2|
\ll |Re\{\lambda_A|$\}), the eigenvalue corresponding to the dark
component has a non-zero negative real component, leading to a
population loss rate:

\begin{equation}
\label{eq:detLoss} -2 Re\{\lambda_D\} \equiv R_L^{(\Delta_2)} = 4
\frac{|\Omega_{13}|^2 |\Omega_{23}|^2}{\Omega^4}\frac{\Delta_2^2
\Gamma_3^{(\mathrm{t})}}{\Omega^2}
\end{equation}

\noindent  The prediction
$P=\exp[-(R_L^{(\gamma_{12})}+R_L^{(\Delta_2)})t]$ is plotted in the
inset of Fig.~\ref{fig:withDetune}(a) and is seen to agree well with
the numerical results (where $R_L^{(\gamma_{12})}$ was determined by
the numerically calculated loss for $\Delta_{23}=0$).
Eq.~(\ref{eq:detLoss}) shows how the field strength, via $\Omega^2$
in the denominator, determines the frequency width of the EIT
feature.

The above analysis indicates that it is only the two-photon detuning
$\Delta_2$ which effects the relative phase of $\ket{1}$ and
$\ket{2}$ and therefore effects the dark state.   EIT will occur in
the presence of a large one-photon detuning $\Delta_{13}$ and the
inset of Figure~\ref{fig:withDetune}(a) shows such a case with
$\Delta_{13}=(2 \pi)~1~$GHz and $\Delta_{23}$ varied about the
two-photon resonance. The presence of a transparency peak is still
clear. The important difference is, because of the large one-photon
detuning $\Delta_{13} \gg \Gamma_3^{(\mathrm{t})}$, the damping of
the absorbing state $-2 Re\{\lambda_{A}\}$ is substantially reduced
and so both the dephasing $R_L^{(\gamma_{12})}$ and detuning
$R_L^{(\Delta_2)}$ loss are reduced.  The analytic model
(\ref{eq:detLoss}) is not valid for large one-photon detunings
$\Delta_{13}$ where the strong damping assumption $-2
Re\{\lambda_{A}\} \gg |\Delta_2|$ does not hold.  As a result, one
sees non-exponential decay in the large one-photon detuning cases
(upper curves of Fig.~\ref{fig:withDetune}(b).

Figure~\ref{fig:withDetune}(b) shows simulations at two-photon
resonance $\Delta_{13} = \Delta_{23}$, varying the one-photon
detuning $\Delta_{13}$.  The population decay is much slower as the
detuning gets larger.   For comparison, we also plot the decay for
initial state equal to the absorbing state
$\ket{\Psi_{\mathrm{init}}}=(\ket{1}+\ket{2})/\sqrt{2}$.  We note
that the analytic model for loss of the dark state
(\ref{eq:detLoss}) is invalid for large one-photon detunings, where
the absorbing state is not completely damped.

\subsection{\label{subsec:measurement} Effective $\Lambda$-system via tunneling and measurement}

Thus far we have considered the system to be a three level system
with the excited level $\ket{3}$ subject to a fast population decay
$\Gamma_3^{(\mathrm{t})}$.  Underlying this decay are actually two
processes: the fast resonant tunneling to a near degenerate level in
the right hand well ($\sigma_{34}$) followed by interwell relaxation
and possibly a strong measurement of the population in $\ket{4}$
($\Gamma_4$); see Fig.~\ref{fig:pcqubit}(c). We show here how the
picture of a three-level system with a strong damping of $\ket{3}$
(the Hamiltonian (\ref{eq:Hideal})) is most valid when $\sigma_{34}
< \Gamma_4$ but actually has a larger range of validity than one
might expect.  We derive an expression for $\Gamma_3^{(\mathrm{t})}$
and also see how the tunneling slightly shifts $\omega_3$.

To do this we consider the Schr\"odinger evolution of the full
four-level system Hamiltonian (with the same frame transformation
as Eq.~(\ref{eq:Hideal}) and $\delta_4=\Delta_{13}+\delta_{34}$,
where $\delta_{34}=\omega_4-\omega_3$):

\begin{align}
 {\cal{H}} = \frac{\hbar}{2} \left[\begin{matrix}
         0 & 0 & \Omega_{13}^* & 0\\
     0 &-2 \Delta_2 & \Omega_{23}^* & 0 \\
      \Omega_{13}& \Omega_{23} & -2 \Delta_{13}^{(0)} & 2 \sigma_{34} \\
      0 & 0 & 2 \sigma_{34} & - i \Gamma_4 -
      2 (\Delta_{13}^{(0)}+\delta_{34})
         \end{matrix}\right],
 \label{eq:Hmeasure}
\end{align}

\noindent We have used the notation $\Delta_{13}^{(0)}$ to
distinguish it from $\Delta_{13}$ which includes the frequency
shift of $\omega_3$ induced by $\ket{4}$.

To recover our three-level picture, we note that when $\Gamma_4 \gg
\sigma_{34}$ we can adiabatically eliminate level $\ket{4}$ to
obtain $c_4=-2 c_3 \sigma_{34}/[2 (\delta_{34}+\Delta_{13}^{(0)}) -
i \Gamma_4]$. Plugging this result back into the equation for
$\dot{c}_3$ reveals that our system can be reduced to a three-level
system as in (\ref{eq:Hideal}) with $\Gamma_3^{\mathrm{(t)}} = 4
|\sigma_{34}|^2 \Gamma_4/[\Gamma_4^2+4
(\delta_{34}+\Delta_{13}^{(0)})^2]$ and
$\Delta_{13}=\Delta_{13}^{(0)}+4|\sigma_{34}|^2\delta_{34}/[\Gamma_4^2+4(\delta_{34}+\Delta_{13}^{(0)})^2]$.
Alternatively, when $\Gamma_4 \ll \sigma_{34}$ we would expect the
tunneling to induce a splitting of $\ket{3}$ and $\ket{4}$ into two
superposition eigenstates (split by $2 \sigma_{34}$).

\begin{figure}
\includegraphics{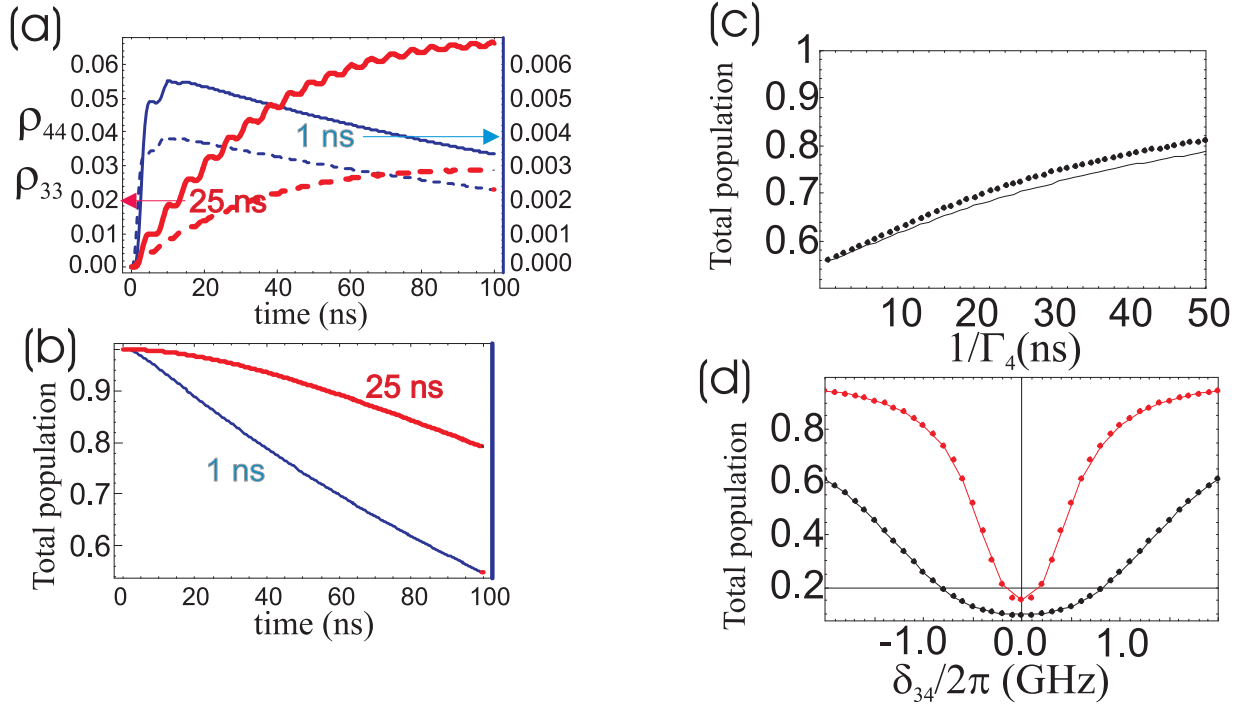}
\caption{\label{fig:measurementTime} (Color online)
\textbf{Consequences of the measurement state characteristics.}
 \textbf{(a)} The thinner curves (blue) show
populations $\rho_{44}$ (solid curves) and $\rho_{33}$ (dashed) as a
function of time for a fast read-out $\Gamma_4^{-1}=1~\mathrm{ns} <
\sigma_{34}^{-1}$ (with scale on the right side). After initial
transient period, the two values reach quasi-steady state values,
which undergo slow exponential decay. Conversely the thick (red
online) curves show a slow measurement case
$\Gamma_4^{-1}=25~\mathrm{ns} \gg \sigma_{34}^{-1}$ (scale on the
left).  In this case the populations do not reach a quasi-steady
over the time scale plotted. \textbf{(b)} The total population
remaining versus time for the same two cases.  \textbf{(c)} The
population remaining at 100~ns for varying measurement rates (dots)
and compared to the prediction of a three-level system with
$\Gamma_3^{(\mathrm{t})}$. As the measurement gets slower, this
prediction slightly and increasingly underestimates the actual
population which should be observed. \textbf{(d)} Numerical results
(dots) and three-level model predictions (solid curves) now letting
$\delta_{34}$ vary, for the cases $\Gamma_4^{-1}=1~\mathrm{ns}$
(black, lower curve) and 10~ns (upper, red curve).}
\end{figure}

We carried out several numerical simulations of the four-level
density matrix equations (\ref{eq:dmEvol}) for this system,
considering first the resonant case ($\delta_{34}=0$).  In them we
used $\sigma_{34}=(2 \pi) 150~\mathrm{MHz}=(1.2~ns)^{-1}$, fields
$\Omega_{13}=(2 \pi)120~$MHz, $\Omega_{23}=(2 \pi)150~$MHz, the
corresponding dark initial state $\rho_{11}=0.61, \rho_{22}=0.39,
\rho_{12}=-\sqrt{\rho_{11} \rho_{22}}$ ({\it i.e.} full coherence),
and a dephasing rate $\gamma_{12}=(2 \pi) 2$~MHz.
Figure~\ref{fig:measurementTime}(a) shows the populations
$\rho_{33},\rho_{44}$ versus time for cases
$\Gamma_4^{-1}=1~\mathrm{ns} = ((2 \pi)159$ MHz)$^{-1}$ (thin, blue
curves) and $25~\mathrm{ns} = ((2 \pi)6$ MHz)$^{-1}$ (thick, red
curves).   Fig.~\ref{fig:measurementTime}(b) shows the total
population remaining versus time.   In the fast measurement case,
1~ns, $\rho_{33}$ and $\rho_{44}$ are seen to track each other, and
we see the exponential decay of the population as in the previous
examples. For the slower measurement case, 25~ns, we see $\rho_{33}$
and $\rho_{44}$ still track each other, but here there is a large
excitation $\rho_{33}$ (note the different scale), no fast
transients in $\rho_{33}$ to a quasi-steady state value, and
non-exponential population decay, all indications of the breakdown
of EIT.

To check the validity of the effective three-level model, in
Fig.~\ref{fig:measurementTime}(c) we compare its predictions for the
populations remaining after 100~ns  (solid curve) with predictions
of the full four-level model (dots). The agreement is excellent for
the $1/\Gamma_4 \le 5$~ns, and is still in rough agreement even up
to 50~ns.  The adiabatic elimination procedure appears to be valid
well beyond the expected regime $\Gamma_4^{-1} \ll
\sigma_{34}^{-1}$. The breakdown of EIT in the 25~ns case is due to
$\Gamma_3^{(t)}$ becoming too large (see
Section~\ref{subsec:dephasing}), rather than a breakdown of the
effective three-level model.

In Fig.~\ref{fig:measurementTime}(d) we show the population
remaining for when $\delta_{34}$ is varied (for both
$1/\Gamma_4=1~$ns (black curves) and 10~ns (red curves)). We have
kept the microwave fields on bare resonance
$\Delta_{13}^{(0)}=\Delta_{23}^{(0)}=0$ and accounted for the
predicted frequency shifts of $\omega_3$ in the three-level model.
As $\delta_{34}$ becomes comparable to $\Gamma_4$ the tunneling rate
$\Gamma_3^{(t)}$ goes down as predicted.

In summary, we find that so long as $\Gamma_4 > \sigma_{34}$ the
simple three-level provides an excellent model and even when
$\Gamma_4 \sim \sigma_{34}$ or somewhat larger, this model
unexpectedly gives very good predictions of the behavior. However,
one must be careful of the strong dependence of $\Gamma_3^{(t)}$ on
$\Gamma_4$, as this may severely effect the necessary conditions for
EIT (which are discussed in Section~\ref{subsec:dephasing}). Through
$\Gamma_4$ the SQUID measurement rate can thus have a large
influence on the EIT. When the detuning $\delta_{34}$ becomes
comparable to $\Gamma_4$ the tunneling rate is reduced as expected
and one must account for the frequency shift of $\omega_3$. For the
remainder of the paper, we will not explicitly include $\ket{4}$ in
the calculations, but assume some $\Gamma_3^{(t)} \sim 1$~ns and
that the frequency shift is already included in the definitions of
$\Delta_{13},\Delta_{23}$.

\section{EIT in the presence of decoherence, incoherent population exchange, and qubit tunneling}
\label{sec:loss}

An outstanding, important issue in the eventual application of
SQCs to quantum computing is the identification and suppression of
sources of decoherence and unwanted dynamics of the qubit states
($\ket{1}$ and $\ket{2}$). In particular, decoherence from pure
dephasing \cite{decoherence}, intrawell relaxation, interwell
resonant tunneling \cite{resCoup}, and coupling to microscopic
degrees of freedom in the junction \cite{microRes} have all been
proposed as potential hurdles in successfully isolating a coherent
two-level system for use as a qubit.  Use of phase sensitive
methods such as EIT could be a fruitful path for exploring and
differentiating the contributions of these various decoherence
processes to qubit dynamics. To learn how EIT is effected by
decoherence, we use the density matrix approach here to include
pure dephasing, population loss, incoherent population exchange,
and resonant coupling to the right well. We find a minimum
microwave coupling strength necessary for the observation of EIT
in the presence of decoherence and see it contributes small
exponential loss (as seen in the numerical results above). We
derive analytic expressions for the loss rates, which are
proportional to the decoherence processes but with coefficients
which depend on the nature of the process.  These results are a
generalization of the results for pure dephasing previously
published \cite{EIT}.

EIT is a unique tool to probe decoherence which compliments the
previously explored techniques of spin echo \cite{spinEcho} and
Rabi oscillation decay \cite{microRes}, and this section will
demonstrate some of its advantages. First, in the limit of small
decoherence, the system is left completely undisturbed by the
probe.  Second, the dependence of the loss rates on relative field
strengths can be used to determine the nature of the decoherence
process. Third, the populations of the qubit states $\ket{1}$
$\ket{2}$ do not need to me manipulated in the process (with $\pi$
pulses, etc.). Besides the advantage of simplicity, this latter
point also leaves open the possibility of decoherence rates which
have a non-trivial dependence on the relative state populations.

\subsection{Density matrix approach}
\label{subsec:bloch}

To carry out this analysis, we must go beyond the Schr\"odinger
approach, and introduce the corresponding density matrix evolution
(\ref{eq:dmEvol}), also referred to as the Bloch equations
\cite{Scully}.  We work in the three-level case and transform to
the frame defined above Eq.~(\ref{eq:Hideal}) and obtain:

\begin{eqnarray}
\label{eq:OBEdephase} \dot{\rho}_{11} & = & - \Gamma_1^{(t)}
\rho_{11} + \Gamma_{2 \rightarrow 1} \rho_{22}
 -\frac{i}{2} \Omega_{13}^* \rho_{31} + \frac{i}{2} \Omega_{13} \rho_{13},   \nonumber  \\
 \dot{\rho}_{22} & = & - (\Gamma_2^{(t)} + \Gamma_{2 \rightarrow 1}) \rho_{22} - \frac{i}{2}
\Omega_{23}^* \rho_{32} + \frac{i}{2} \Omega_{23} \rho_{23},  \nonumber \\
\dot{\rho}_{33} & = & - \Gamma_3^{(\mathrm{t})} \rho_{33} +
\frac{i}{2} \Omega_{13}^* \rho_{31} - \frac{i}{2} \Omega_{13}
\rho_{13} + \frac{i}{2}
\Omega_{23}^* \rho_{32} - \frac{i}{2} \Omega_{23} \rho_{23},   \nonumber \\
\dot{\rho}_{12} & = &
-\big(\gamma_{12}+\frac{\Gamma_1^{(t)}+\Gamma_2^{(t)} + \Gamma_{2
\rightarrow 1}}{2}\big)\rho_{12}-\frac{i}{2}\Omega_{13}^*\rho_{32}
+\frac{i}{2}\Omega_{23}\rho_{13},    \nonumber \\
\dot{\rho}_{13} & = &  -
\frac{\Gamma_3^{(t)}+\Gamma_1^{(t)}}{2}\rho_{13}+\frac{i}{2}\Omega_{13}^*(\rho_{11}-\rho_{33})
+\frac{i}{2}\Omega_{23}^*\rho_{12},   \nonumber  \\
\dot{\rho}_{23} & = &  -
\frac{\Gamma_3^{(t)}+\Gamma_2^{(t)}+\Gamma_{2 \rightarrow 1}}{2}
\rho_{23}+\frac{i}{2}\Omega_{23}^*(\rho_{22}-\rho_{33})
+\frac{i}{2}\Omega_{13}^*\rho_{21}.
\end{eqnarray}

\noindent For simplicity, we have supposed the resonant case
$\Delta_{13}=\Delta_{23}=0$ and ignored inter-well relaxation of
$\ket{3}$, which is dominated by $\Gamma_3^{(t)}$. The remaining
three elements equations are determined by
$\rho_{ij}^*=\rho_{ji}$. The most important new piece here is the
decoherence rate of $\rho_{12}$:
$\gamma_{12}+(\Gamma_1^{(t)}+\Gamma_2^{(t)} + \Gamma_{2
\rightarrow 1})/2$.  This will decohere the dark state and lead to
the small losses in the population.

To proceed, we adiabatically eliminate the excited state
coherencess $\rho_{13},\rho_{23}$ as they are strongly damped by
$\Gamma_3^{(t)}$. In these equations we can ignore
$\Gamma_1^{(t)},\Gamma_2^{(t)} \ll \Gamma_3^{(t)}$ as well as
$\rho_{33} \ll \rho_{11},\rho_{22}$. We then plug the results back
into the remaining equations to obtain:

\begin{eqnarray}
\label{eq:OBEadElim} \dot{\rho}_{11} & = & - \Gamma_1^{(t)}
\rho_{11} + \Gamma_{2 \rightarrow 1} \rho_{22}
 -\frac{|\Omega_{13}|^2}{\Gamma_3} \rho_{11} - \bigg(\frac{\Omega_{13}\Omega_{23}^*}{2 \Gamma_3} \rho_{12} + \mathrm{c.c.}),   \nonumber  \\
 \dot{\rho}_{22} & = & - (\Gamma_2^{(t)}
+ \Gamma_{2 \rightarrow 1}) \rho_{22}
 -\frac{|\Omega_{23}|^2}{\Gamma_3} \rho_{22} - \bigg(\frac{\Omega_{13}\Omega_{23}^*}{2 \Gamma_3} \rho_{12} + \mathrm{c.c.}),   \nonumber  \\
\dot{\rho}_{12} & = &
-\big(\gamma_{12}+\frac{\Gamma_1^{(t)}+\Gamma_2^{(t)} + \Gamma_{2
\rightarrow 1}}{2}\big)\rho_{12}-\frac{\Omega^2}{2
\Gamma_3^{(t)}}\rho_{12}
-\frac{\Omega_{13}^*\Omega_{23}}{2\Gamma_3^{(t)}}(\rho_{11}+\rho_{22})
\end{eqnarray}

\noindent We note here a strong damping of the coherence provided
by the fields $\Omega^2/2 \Gamma_3^{(t)}$.  This damping acts to
drive the system into the dark state. In the limit that the
decoherence terms ($\gamma_{12}, \, \Gamma_1^{(t)}, \,
\Gamma_2^{(t)}, \, \Gamma_{2 \rightarrow 1}$) vanish, there is a
steady state solution consisting of perfect coherence:
$\rho_{11}=|\Omega_{23}|^2/\Omega^2$,
$\rho_{22}=|\Omega_{13}|^2/\Omega^2$, $\rho_{12} = -\Omega_{13}
\Omega_{23}/\Omega^2$.  However, the decoherence terms drive the
system out of the dark state, causing excitation $\rho_{33}$. One
sees the ratio of the decoherence rate compared with the EIT
preparation rate $\Omega^2/\Gamma_3^{(t)}$ determines the degree
to which the coherence deviates from the perfect dark state value.

\subsection{Measuring dephasing with EIT}
\label{subsec:dephasing}

\begin{figure}
\includegraphics{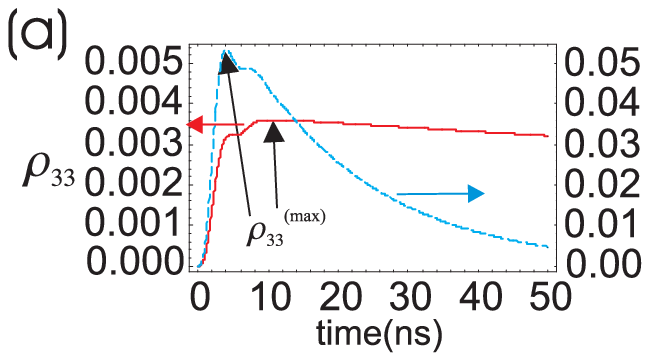}
\includegraphics[width=2in]{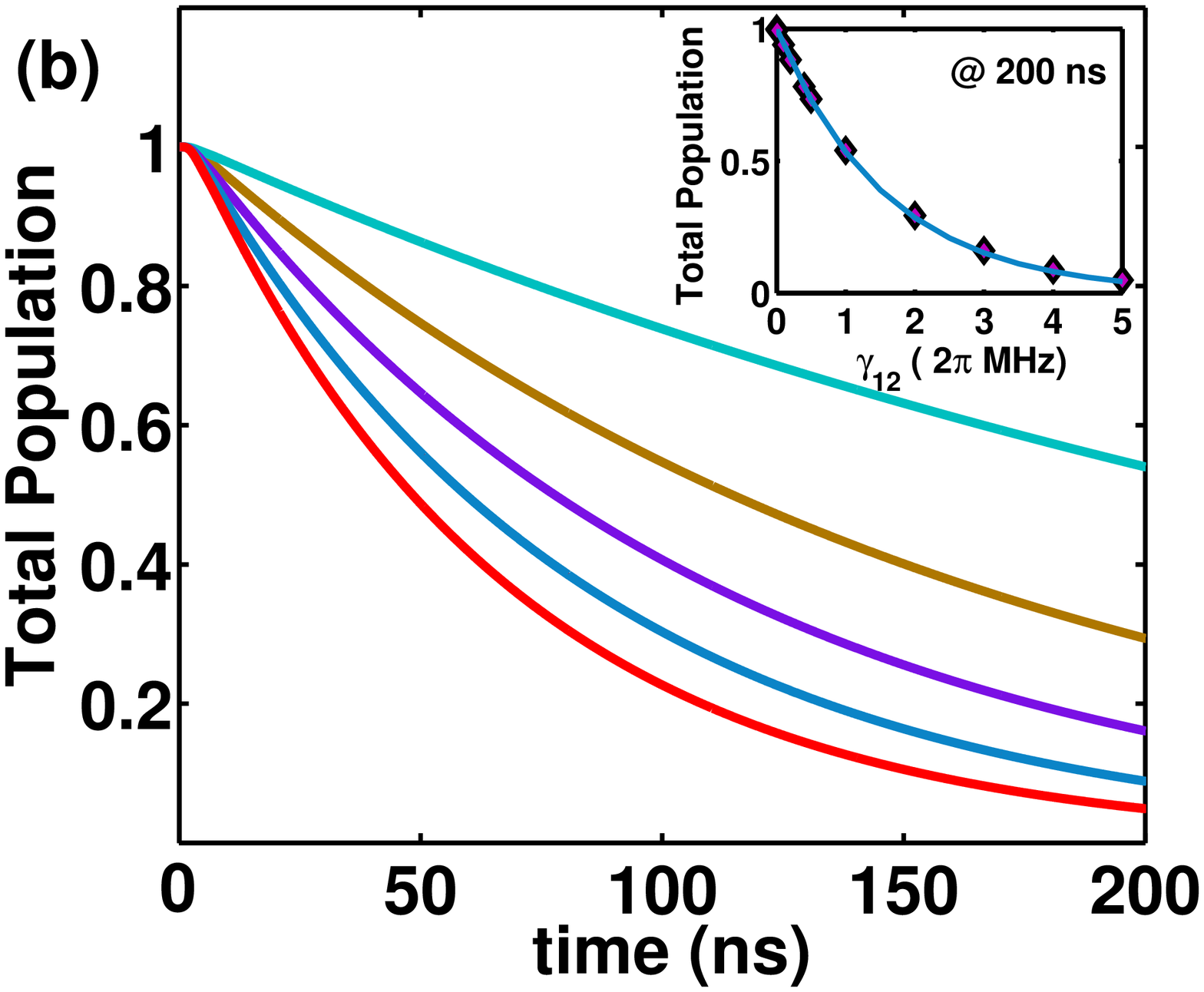}
\includegraphics{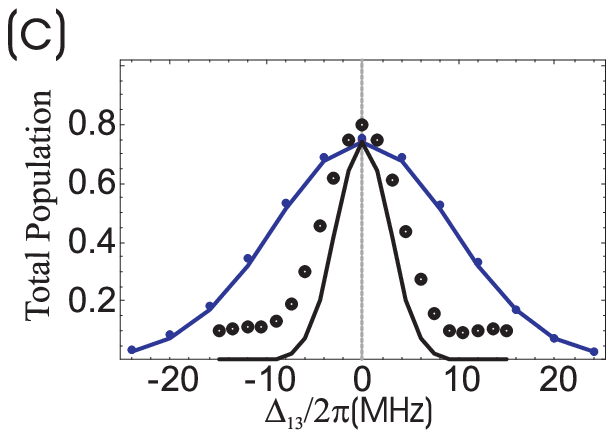}
\caption{\label{fig:dephasing} (Color online) \textbf{EIT loss due
to pure dephasing} \textbf{(a)} The population $\rho_{33}$ versus
time in the presence of a pure dephasing $\gamma_{12}=(2 \pi)1~$MHz
(solid, red curve, scale on left) and $\gamma_{12}=(2 \pi)20~$MHz
(dashed, blue curve, scale on right). In the slow dephasing case
$\rho_{33}$ quickly reaches a plateau value
$\rho_{33}^{\mathrm{(max)}}$ then undergoes a slow exponential
decay. For the faster dephasing, the exponential decay time is
similar to the time required to reach the maximum value.  In each
case, the initial state is $\rho_{11}=\rho_{22}=0.5$ and
$\rho_{12}=-0.5$, $\Omega_{13}=\Omega_{23}=(2 \pi) 150$~MHz and
$\Gamma_3^{(t)}=(2 \pi)130$~MHz \textbf{(b)} The population decay
with varying dephasing rates (top curve to bottom curve)
$\gamma_{12}=(2 \pi)~$ 1, 2, 3, 4, and 5~MHz.  Inset: The population
in the left well at 200~ns versus $\gamma_{12}$ (dots). The solid
curve shows the analytic prediction $\exp(-R_L^{(\gamma_{12})} t)$
based on Eq.~\ref{eq:lossRate}, demonstrating how the population
loss can be used as a probe of $\gamma_{12}$. \textbf{(c)}
Population remaining at 100~ns versus detuning $\Delta_{13}$
(keeping $\Delta_{23}=0)$ for two different field intensities. In
these simulations we used the initial conditions $\rho_{11}=0.61, \,
\rho_{22}=0.39, \, \rho_{12}=-\sqrt{\rho_{11}\rho_{22}}$, the fields
$\Omega_{13}=0.8\,\Omega_{23}$, $\Gamma_3^{(t)}=(2 \pi)159$~MHz,
$\gamma_{12}=(2 \pi)1$~MHz.  The solid curves are the analytic
prediction described in the text for
 $\Omega_{23}=(2 \pi) 100$~MHz (blue, upper curve), and 30~MHz (black, lower curve).
 The dots show the numerical results.}
\end{figure}

We first show this comes into play for the pure dephasing, which is
expected to be the case in many practical implementations and was
analyzed previously in \cite{EIT}. Figure~\ref{fig:dephasing}(a)
(solid, red curve) shows the excited state population $\rho_{33}$
when $\gamma_{12}=(2 \pi) 1~\mathrm{MHz}$ and we apply the fields
$\Omega_{13}=\Omega_{23}=(2 \pi) 150~\mathrm{MHz}$ to the dark state
$\ket{\Psi_\mathrm{init}}=\ket{\Psi_D}=(\ket{1} -
\ket{2})/\sqrt{2}$. One sees a small (note the scale in
Fig.~\ref{fig:dephasing}(a)) but finite excitation. In particular,
there is an initial fast transient behavior to some plateau value
(over a time scale determined by
$\mathrm{Min}\{\Omega^2/\Gamma_3^{(t)}, \Gamma_3^{(t)}\}$), followed
by a slow exponential decay.   The general behavior of an initial
transient rise, Fig.~\ref{fig:dephasing}(a), was seen over a wide
parameter regime. This quasi-steady state excitation of $\ket{3}$ is
the origin of the exponential losses at rate $R_L^{(\gamma_{12})}$
in the previous simulations. Fig.~\ref{fig:dephasing}(b) shows the
exponential decay for several different dephasing rates
$\gamma_{12}$. The inset of Fig.~\ref{fig:dephasing}(b) plots the
populations remaining at 200~ns, which is seen to approach unity as
$\gamma_{12} \rightarrow 0$.

The loss rate can be quantified by considering the $3 \times 3$
evolution matrix for $\rho_{11}$, $\rho_{22}$, and $\rho_{12}$
defined by (\ref{eq:OBEadElim}). Looking at the eigenvalue
corresponding to the smallest loss rate, and expanding to first
order in $\gamma_{12}$ gives:

\begin{eqnarray}
\label{eq:lossRate} R_L^{(\gamma_{12})}  = 2 \gamma_{12}
\frac{|\Omega_{13}|^2 |\Omega_{23}|^2}{\Omega^4}
\end{eqnarray}

\noindent By measuring this decay constant experimentally, one can
use (\ref{eq:lossRate}) to extract $\gamma_{12}$.   Note that the
two rates are simply related by a constant or order unity,
determined by the relative strength of the two microwave field
couplings (in the example in Fig.~\ref{fig:dephasing}, the constant
is $2 \Omega_{13}^2 \Omega_{23}^2/\Omega^4=0.5$). A glance at
Fig.~\ref{fig:dephasing}(b) reveals how choosing the observation
time $t \sim \gamma_{12}^{-1}$ will give the best sensitivity in the
measurement.

The inset of Fig.~\ref{fig:dephasing}(b) shows the prediction
(\ref{eq:lossRate}) in comparison with the analytic results and we
see good agreement.  The adiabatic elimination and the expansion
for small $\gamma_{12}$ require $2 \gamma_{12}
\Gamma_3^{(t)}/\Omega^2 \ll 1$ and $\gamma_{12} \ll
\Gamma_3^{(t)}$.  This ratio is 0.07 for $\gamma_{12}=(2
\pi)~5$~MHz.  For higher dephasing rates, the dephasing rate
competes with the preparation rate, and the first order expansion
in $\gamma_{12}$ becomes less valid. Such a case is seen in
Fig.~\ref{fig:dephasing}(a) (dashed, blue curve) where we plot
$\rho_{33}$ for a case with $\gamma_{12}=(2 \pi)~20$~MHz. The
exponential decay occurs with a time scale comparable to the
transient time to reach the quasi-steady state plateau. In such
cases, $\gamma_{12}$ can only be estimated from the tunneling rate
by more detailed modeling of the underlying Bloch equations
(\ref{eq:OBEdephase}).  We note that the microwave field intensity
can be adjusted to control $\Omega^2$ and to bring us into a
regime where (\ref{eq:lossRate}) is valid. The breakdown of this
inequality occured in the slow measurement time (25~ns) case
plotted in Fig.~\ref{fig:measurementTime}(a,b), for which
$\Gamma_3^{(t)}\approx 4 |\sigma_{34}|^2/\Gamma_4$ became quite
large ($\sim$~GHz).

When detuning and dephasing are both present but sufficiently small,
the two effects add linearly.  In Fig.~\ref{fig:dephasing}(c) we
show the population remaining versus detuning $\Delta_{13}$, with
$\gamma_{12}=(2 \pi)1$~MHz (with $\Delta_{23}=0$) both in a strong
field ($2 \gamma_{12} \Gamma^{(t)}/\Omega^2=0.016$) and weak field
($2 \gamma_{12} \Gamma^{(t)}/\Omega^2=0.18$) case. The prediction
$P=\exp(-(R_L^{(\gamma_{12})}+R_L^{(\Delta_2)})t)$ holds for the
stronger field but overestimates the loss for the weaker field.

We conclude from the above calculations that there are two basic
conditions that have be satisfied for a reliable measurement of
the decoherence in the system.  First, the decoherence rate
$\gamma_{12}$ should be much smaller than the loss rate
$\Gamma_3^{(\mathrm{t})}$, which holds in systems of interest.
Second, we must be able to apply sufficiently strong fields that
the preparation rate $\Omega^2/\Gamma_3^{(\mathrm{t})}$ dominates
$\gamma_{12}$.

\subsection{EIT with incoherent population loss and exchange}
\label{subsec:incoherent}

When the decoherence of $\rho_{12}$ occurs due to population loss
and exchange instead of dephasing, the effect on the EIT is much
the same, with the $\gamma_{12}$ simply replaced by the total
decoherence rate.  However, because these processes involve
additional changes in the populations, the population loss rate,
which we use to diagnose the decoherence rate, will be different.

\begin{figure}
\includegraphics{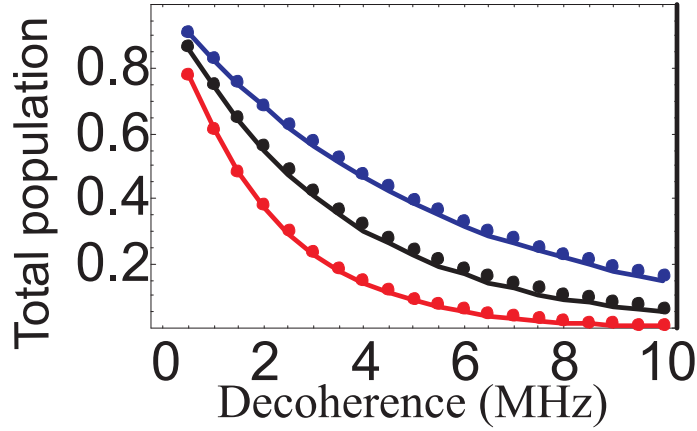}
\caption{\label{fig:lossTypes} (Color online) \textbf{EIT loss with
dephasing, open loss, and closed loss} The population loss after
100~ns for several types of decoherence.  The parameters are as in
Fig.~\ref{fig:dephasing}(c) but with $\Omega_{23}=(2 \pi)150$~MHz.
The curves show the analytic predictions (\ref{eq:lossRate}) and
(\ref{eq:lossRatePop}) and the dots show numerical results. The
lowest curve is for purely open loss $\Gamma_2^{(t)}$, the middle
curve for pure dephasing $\gamma_{12}$, and the top curve shows pure
population exchange $\Gamma_{2 \rightarrow 1}$.  The horizontal axis
shows the corresponding decoherence rate for each case:
$\Gamma_2^{(t)}/2,\,\gamma_{12}, \, \Gamma_{2 \rightarrow 1}/2$,
respectively.}
\end{figure}

Referring back to evolution matrix  (\ref{eq:OBEadElim}), we again
find the eigenvalues to determine the loss rate of the dark state.
Expanding to first order in
$\Gamma_1^{(t)},\Gamma_2^{(t)},\Gamma_{2 \rightarrow 1}$,
respectively, we find:

\begin{eqnarray}
\label{eq:lossRatePop} R_L^{(\Gamma_1^{(t)})} & = & \Gamma_1^{(t)}
\frac{|\Omega_{23}|^2}{\Omega^2}; \nonumber \\
R_L^{(\Gamma_2^{(t)})} & = &\Gamma_2^{(t)}
\frac{|\Omega_{13}|^2}{\Omega^2}; \nonumber \\
R_L^{(\Gamma_{2 \rightarrow 1})} &= &\Gamma_{2 \rightarrow 1}
\frac{|\Omega_{13}|^4}{\Omega^4}
\end{eqnarray}

\noindent It is interesting to note that the coefficient will
depend in different ways on the relative intensities of the two
fields depending on the origin of the decoherence.
Fig.~\ref{fig:lossTypes} shows the population loss after 100~ns
for different kinds of loss, each plotted versus the total
decoherence rate of $\rho_{12}$.  The open system loss
$\Gamma_2^{(t)}$ is greater than the pure dephasing case because
there is direct population loss on top of the absorption into
$\rho_{33}$ due to decay out of the dark state.  The closed system
loss (intrawell relaxation $\Gamma_{2 \leftarrow 1}$) is seen to
be smaller than pure dephasing, however, (\ref{eq:lossRatePop})
shows that this could be greater or smaller depending on the
relative values of $\Omega_{13}$ and $\Omega_{23}$.

While we assumed in this discussion that the intra-well relaxation
of $\ket{3}$ was negligible ($\Gamma_{3 \rightarrow 1}+\Gamma_{3
\rightarrow 2} \ll \Gamma_3^{(t)}$), the present discussion is
easily generalized when this can not be assumed.  In this case the
population loss rate is the same as before, multiplied by a factor
reflecting the proportion of atoms in $\ket{3}$ which actually
tunnels to the right well: $\Gamma_3^{(t)}/(\Gamma_{3 \rightarrow
1}+\Gamma_{3 \rightarrow 2})$.  This reflects the fact that
excitation of $\ket{3}$ due to decoherence will only be registered
as population loss upon tunneling to the right well.

\subsection{\label{subsec:tunneling} Resonant tunneling loss out of left well}
\label{subsec:tunelling}

Just as the decay of $\ket{3}$ is the result of tunneling followed
by decay of $\ket{4}$, interwell tunneling of $\ket{2}$, which is
conceivably a leading order effect in the decoherence, involves
tunneling to a near resonant level $\ket{5}$ (see
Fig.~\ref{fig:pcqubit}(c)).  Here we consider this effect in detail
to find the conditions where an effective damping rate
$\Gamma_2^{(t)}$ can be used, and also explore conditions where the
dynamics are more complicated.

To explore this issue, we have performed numerical simulations for
the system $\{\ket{1},\ket{2},\ket{3},\ket{5}\}$ where level
$\ket{2}$ is detuned from $\ket{5}$ by $\delta_{25} \equiv
\omega_5-\omega_2$. The Hamiltonian (with the transformation
$\delta_5 = \Delta_2+\delta_{25}$) is:

\begin{align}
 \mathcal{H} = \frac{\hbar}{2} \left[\begin{matrix}
         0 & 0 & \Omega_{13}^* & 0\\
     0 & -2 \Delta_2& \Omega_{23}^* & 2 \sigma_{25} \\
      \Omega_{13}& \Omega_{23} & - 2 \Delta_{13} -  i \Gamma_3^{(\mathrm{t})}  & 0 \\
      0 & 2 \sigma_{25}  & 0 & -  i \Gamma_5
     -2 (\Delta_2+ \delta_{25}))
         \end{matrix}\right],
 \label{eq:HtunnelLoss}
\end{align}

\noindent Mathematically similar energy level structures have been
considered in the context of atomic systems \cite{fourlevelRes}.

\begin{figure}
\includegraphics{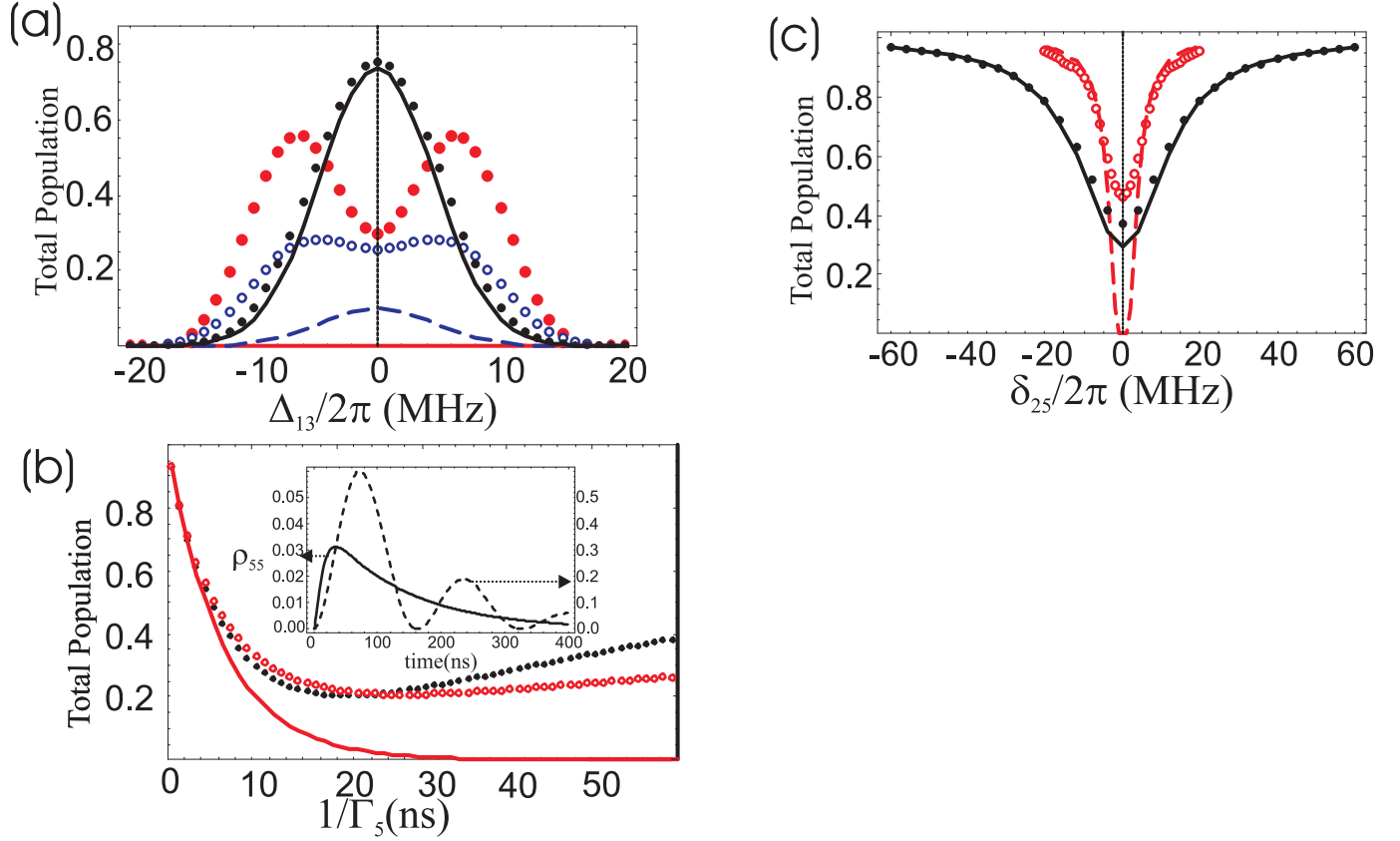}
\caption{ \label{fig:tunnelingLoss} (Color online) \textbf{Loss due
to resonant tunneling to right well.} In the simulations we assume a
(resonant) tunneling rate $\sigma_{25}=(2 \pi)5$~MHz, with fields
$\Omega_{13}=0.8 \, \Omega_{23}$ and the corresponding dark state
$\rho_{11}=0.61, \, \rho_{22}=0.39$, $\rho_{12}=-\sqrt{\rho_{11}
\rho_{22}}$. \textbf{(a)}  The population remaining at 100~ns of
fields applied with strength $\Omega_{23}=50$~MHz, versus the
detuning $\Delta_{13}$ (keeping $\Delta_{23}=0$), but varing the
relaxation time $\Gamma_5^{-1}$. For the small solid dots (black)
$\Gamma_5^{-1}=2$~ns.   The solid curve shows the analytic
prediction based on the effective loss rate $\Gamma_2^{(t)}$
described in the text.  The open (blue) dots show a case
$\Gamma_5^{-1}=15$~ns, in which case a splitting appears in the
resonance, contrary to the analytic prediction (dashed curve).  The
large (red) dots show the case $\Gamma_5^{-1}=80$~ns, for which the
splitting becomes more pronounced and the loss rate quite small,
while the $\Gamma_2^{(t)}$ model predicts complete loss of the
population. \textbf{(b)}  The population remaining at 100~ns at the
two photon resonance ($\Delta_{13}=\Delta_{23}=0$) versus
$\Gamma_5^{-1}$. The solid (black) dots show the case
$\Omega_{13}=(2 \pi)150$~MHz and the open (red) dots show
$\Omega_{13}=(2 \pi)50$~MHz.  They roughly agree with each other and
the $\Gamma_2^{(t)}$ model (solid curve) for $\Gamma_5^{-1} \ll
\sigma_{25}^{-1}=32$~ns. However, for larger $\Gamma_5^{-1}$ the
loss becomes slower.  The inset shows the population $\rho_{55}$ for
the a fast (5~ns, solid curve) and slow (80~ns, dashed) relaxation
times $\Gamma_5^{-1}$ (note the different scales). The fast case
looks analogous to decoherence (see Fig.~\ref{fig:dephasing}(a)),
while oscillations occur in the slow case. \textbf{(c)} The
population remaining versus the level detuning $\delta_{25}$ in the
case $\Gamma_5^{-1}=$8~ns (solid, black dots) and 80~ns (open, red
dots). The solid and dashed curves show the
$\Gamma_2^{\mathrm{(t)}}$ model predictions. }
\end{figure}

In the following we take $\sigma_{25} = (2 \pi) 5$~MHz and set the
pure dephasing $\gamma_{12}=0$ to isolate the contribution from
the presently considered effect.  Analogous to
Section~\ref{subsec:measurement}, when $\Gamma_5 \gg \sigma_{25}$
one can easily reduce the system to an effective-three level
system with an additional loss rate $\Gamma_2^{\mathrm{(t)}} = 4
|\sigma_{25}|^2 \Gamma_5/(\Gamma_5^2+4 \delta_{25}^2)$.   The
small (black) dots in Fig.~\ref{fig:tunnelingLoss}(a) present the
population as a function of the detuning $\Delta_{13}$ (keeping
$\Delta_{23}=0$) after 100~ns of evolution for a case with
$\Gamma_5^{-1}=2~\mathrm{ns}\approx(15 \, \sigma_{25})^{-1}$ (and
$\delta_{25}=0$). The results are in good agreement with the
prediction one obtains from the loss rate
Eq.~(\ref{eq:lossRatePop}) with this predicted tunneling rate
$\Gamma_2^{(t)}$ (solid curve). Fig.~\ref{fig:tunnelingLoss}(b)
shows the computed loss (dots) for resonant fields
$(\Delta_{13}=0)$ compared with the loss expected from the
calculated $\Gamma_2^{\mathrm{(t)}}$ (solid curve) as a function
of $\Gamma_5^{-1}$, for two different field strengths. One sees
the estimate is good for $\Gamma_5^{-1} < 5~\mathrm{ns}$. The
inset shows $\rho_{55}$ versus time when $\Gamma_5^{-1}=5$~ns. One
sees it quickly reaches a quasi-steady state plateau, then
undergoes an exponential decay.

For larger $\Gamma_5^{-1}$, Fig.~\ref{fig:tunnelingLoss}(b) shows
the loss begins to decrease in contrast to the analytic estimate.
The red curve in Fig.~\ref{fig:tunnelingLoss}(a) shows the
population remaining versus detuning $\Delta_{13}$ in a case in for
$\Gamma_5^{_1}=80$~ns. Besides the $\Gamma_2^{(t)}$ model
incorrectly predicting complete loss of the population after 100~ns,
in the numerical results there is the clear appearance of
double-peaked structure. This can be understood from the coupling
$\sigma_{25}$ giving rise to two eigenstates $(\ket{2}\pm
\ket{5})/\sqrt{2}$ split by $2 \sigma_{25}$, each of which gives
rise to a distinct EIT resonance.  The initial state (no population
in $\ket{5}$) is a superposition of these eigenstates and we get
oscillations of the population between $\ket{2}$ and $\ket{5}$. The
inset in Fig.~\ref{fig:tunnelingLoss}(b) (dashed curve) shows these
oscillations in $\rho_{55}$ for $\Gamma_5^{-1}=80$~ns.  Because of
the weak damping, the oscillations persist and the quasi-steady
state is not reached during the time scale plotted.  The dotted
curve and open dots in Fig.~\ref{fig:tunnelingLoss}(b) show an
intermediate case $\Gamma_5^{-1}=15~\mathrm{ns}$ where the double
peak structure is just becoming apparent, and the analytic estimate
has begun to break down.

In Fig.~\ref{fig:tunnelingLoss}(c) we address the case where the
tunneling levels $\ket{2}$ and $\ket{5}$ can be slightly
off-resonant $(\delta_{25}\not=0)$.  The filled dots show the
population remaining for $\Gamma_5^{-1}=8$~ns versus
$\delta_{25}$. The solid curve shows the
$\Gamma_{2}^{\mathrm{(t)}}$ model estimate, which correctly
accounts for the slower tunneling rate as we move off resonance.
The red dots show the same for $\Gamma_5^{-1}=80$~ns.  In this
limit, the analytic estimate expression severely overestimates the
loss for $\delta_{25} < \Gamma_5$ but as we move off resonance,
the coherent tunneling plays less of a role and the effective
tunneling decay rate model becomes more accurate.

In summary, tunneling of our lower states will be a source of loss
in EIT.  The behavior will depend qualitatively on the relative
strength of the coherent coupling and the loss rate of the
additional quantum level and so can provide us with information
about these quantities.  In the limit where the loss rate dominates,
we see how the it reduces to an open system loss of $\ket{2}$ where
as in the other limit we see a qualitative signature (the splitting
of the resonance) of coherent coupling to another level.

\section{\label{sec:crosstalk} EIT with radiation cross-talk}

To now, we have considered how EIT is affected by decoherence and
tunneling to other levels. Another important consideration to
include is that all levels are dipole coupled and so in principle
coupled by the microwave fields. The RWA allows us to neglect the
majority of the couplings as the dynamics are dominated by
couplings which are near resonant. For example, one need not
consider the coupling of field $b$ on the $\ket{1} \leftrightarrow
\ket{3}$ transition or field $a$ on $\ket{2} \leftrightarrow
\ket{3}$. However, in SQCs, the relative scale of the Rabi
frequencies ($\sim 100$~MHz) to the level spacings ($\sim$~GHz) is
somewhat larger than in typical atomic systems. Thus, it is
important to know the magnitude and type of effects that these
``cross'' couplings can have.  Here we consider, separately, a
case with cross coupling within the three level system, and a case
where fields couple to an additional excited level. In general, we
find these effects can be characterized analytically in terms of
additional loss rates and AC Stark shifts. If one neglected these
effects, there are configurations where one may mistakenly
attribute a loss rate to a dephasing when in fact it is due to
off-resonant field coupling. While we found it was beneficial to
turn the microwave coupling strengths $\Omega^2$ up to overcome
decoherence and detuning, we will see how this can increase the
importance of these cross-talk effects. We will also see how
proper understanding of the effects can allow us be mitigate them
by properly compensating for the Stark shifts. To isolate these
cross-talk effects, we will set other losses and dephasings to
zero in the following.

\subsection{\label{subsec:3LevelCross} Radiation cross-talk in a three-level system}

\begin{figure}
\includegraphics{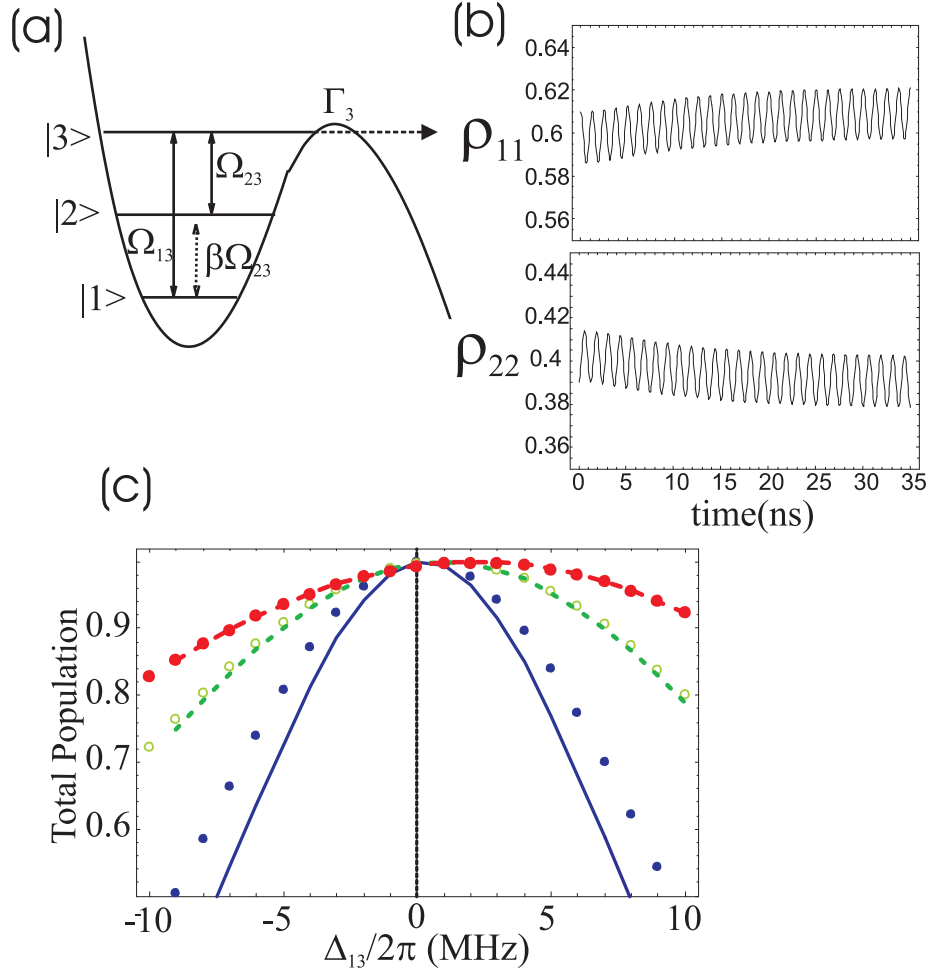}
\caption{\label{fig:crossTalk3levelLadder} (Color online)
\textbf{Cross talk in a ladder system.} \textbf{(a)} Schematic of
the dominant cross-talk term: field $b$ (resonant with
$\ket{2}\leftrightarrow \ket{3}$) also couples
$\ket{1}\leftrightarrow \ket{2}$, detuned by 0.7~GHz. \textbf{(b)}
This induces fast oscillations of the ground state populations
$\rho_{11}$ and $\rho_{22}$ (and a slow overall drift) as shown here
for the initial state $\rho_{11}=0.61, \, \rho_{22}=0.39$ with full
coherence and $\Omega_{13}=0.8 \, \Omega_{23}=(2 \pi) 120~ $MHz and
$\Gamma_3^{(t)}=(2 \pi)159$~MHz. \textbf{(c)} Population remaining
at 50~ns for $\Omega_{23}=(2 \pi) 50$~MHz (small, blue dots),
100~MHz (open, green dots), and 150~MHz (large, red dots), with the
curves showing the analytic predictions based on the AC Stark shifts
described in the text. }
\end{figure}

In the configuration proposed here and in \cite{EIT}, the qubit
states $\ket{1}$ and $\ket{2}$ are the first two levels of a
slightly anharmonic potential (the left well), while the excited
level $\ket{3}$ is the third such level. Therefore, the level
spacing $\omega_3-\omega_2$ is only slightly different than
$\omega_2-\omega_1$ (with the parameters proposed the spacings are
$\sim$30~GHz and the difference is 0.7~GHz).  As a result, the field
$\Omega_{23}$ is only 0.7~GHz detuned from $\ket{1} \leftrightarrow
\ket{2}$ (see Fig.~\ref{fig:crossTalk3levelLadder}(a)). A rough
estimate of this effect was noted in \cite{EIT}. Here we present an
analytic treatment which shows it causes an AC Stark shift which
depends on the relative dipole coupling strengths, field
intensities, and level spacings.   Thus the EIT resonance position
can be a function of field amplitudes used, which can be compensated
by adjusting the field frequencies.

We consider the Hamiltonian (\ref{eq:Hideal}) but do not invoke the
RWA with respect to terms rotating by the mismatch frequency between
the $\ket{2} \leftrightarrow \ket{3}$ and  $\ket{1} \leftrightarrow
\ket{2}$ transitions $\delta \equiv
(\omega_2-\omega_1)-(\omega_3-\omega_2)$.

\begin{align}
 {\cal{H}} = \frac{\hbar}{2} \left[\begin{matrix}
         0 & \beta^* \Omega_{23}^* e^{i \delta t} & \Omega_{13}^*\\
     \beta \Omega_{23} e^{-i \delta t} & -2 \Delta_2 & \Omega_{23}^*  \\
      \Omega_{13}& \Omega_{23} & -2 \Delta_{13} - i \Gamma_3^{(\mathrm{t})} &
         \end{matrix}\right],
 \label{eq:Hcross3}
\end{align}

\noindent where $\beta \equiv x_{12}/x_{23}$ is the ratio of
dipole moments between the additional off-resonant transition and
the intended resonant transition for the field $b$.  In the case
we are considering $\beta=-2.55$ though it should be emphasized
that these ratios are strong functions of the parameters and can
vary by an order of magnitude.

We performed a numerical propagation of the density matrix
equations (\ref{eq:dmEvol}) for this Hamiltonian for the resonant
case $\Delta_{13}=\Delta_{23}=0$ and plot the evolution of
$\rho_{11}$ and $\rho_{22}$ in
Fig.~\ref{fig:crossTalk3levelLadder}(b).  We see the extra
coupling gives rise to a small amplitude, rapid oscillations of
both quantities.  Considering a toy two-level model with only the
off-resonant coupling present predicts population oscillations of
period $(2\pi)/\delta$ and amplitude $\sim |\beta \Omega_{23}|/2
\delta$, in agreement with the numerical results.  The small
deviations of $\rho_{11},\rho_{22}$ from their dark state values
gives rise to absorption into $\ket{3}$ and thus loss.  In the toy
model, an off-resonant coupling can be accounted for as an AC
Stark shift. In particular, $\omega_1$ and $\omega_2$ are
predicted to shift by $\pm \sim |\beta \Omega_{23}|^2/4 \delta$,
respectively. This results in an effective shift of the two-photon
detuning $\Delta_2$ which can be compensated.

Stated in terms of our exponential loss language, the loss rate
$R_L^{(\Delta_2)}$ (\ref{eq:detLoss}) is still valid but the
two-photon detuning $\Delta_2$ should be replaced by
$\Delta_2+\Delta_{AC}^{(12)}$, where

\begin{equation}
\label{eq:ACladder} \Delta_{AC}^{(12)} = \frac{|\beta
\Omega_{23}|^2}{2 \delta}
\end{equation}

\noindent In Fig.~\ref{fig:crossTalk3levelLadder}(c) we plot the
population remaining versus $\Delta_{13}$ (keeping
$\Delta_{23}=0$) for three different values of field intensities.
The solid curves show the theoretical prediction based on the
predicted AC Stark shift. They are in good agreement (the
overestimate of loss at the lowest intensities with some detuning
is due to the damping of the absorbing that being too weak to
efficiently keep the SQC in the dark state). Importantly, if one
adjusts the field frequencies, one can completely avoid loss due
to the cross-talk coupling. Strictly speaking, there is a small
loss if $\ket{2}$ decays at some small rate $\Gamma_2^{(t)}$,
however, this loss is much smaller than the loss already predicted
from the associated decoherence (\ref{eq:lossRatePop}).

\subsection{\label{subsec:4LevelCross} Effect of off-resonant radiation coupling to an additional excited level}

\begin{figure}
\includegraphics{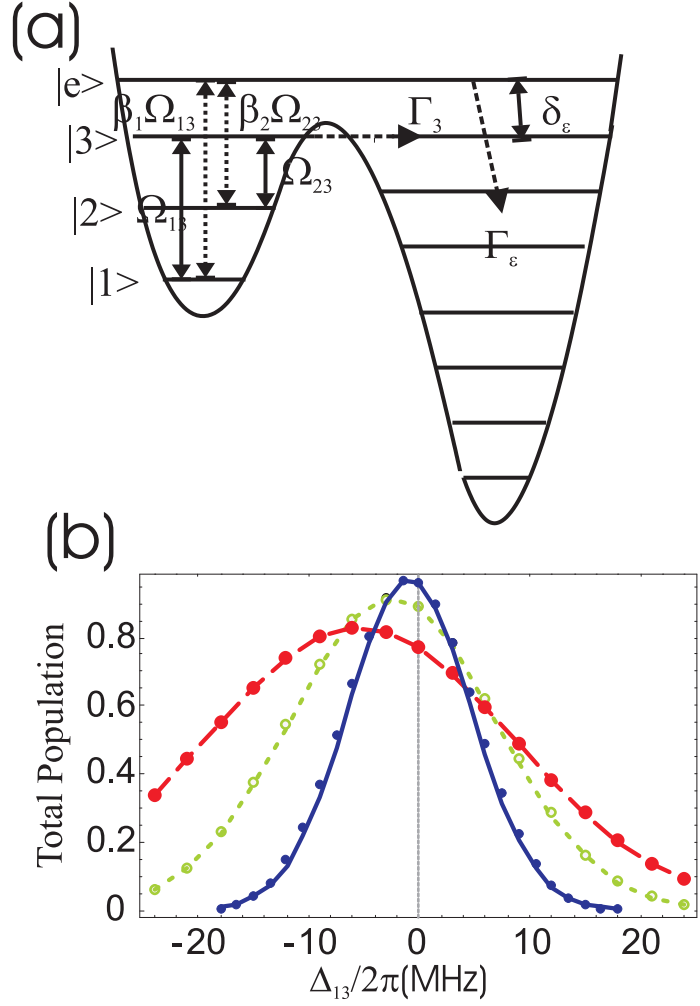}
\caption{\label{fig:crossTalk4level} (Color online) \textbf{Coupling
to an additional excited level. } \textbf{(a)} Schematic of
off-resonant microwave coupling of each of states $\ket{1}$ and
$\ket{2}$ to an additional level $\ket{e}$ above the barrier.
\textbf{(b)} The population remaining at 100~ns for the same initial
state and relative field strengths as in
Fig.~\ref{fig:crossTalk3levelLadder}. We show the cases
$\Omega_{23}=(2\pi)60$~MHz (small solid, blue), 100~MHz (open,
green), and 150~MHz (large solid, red).  The solid curves show the
loss and AC Stark shift predicted in the text (\ref{eq:AClossE}). To
isolate and clearly show the effect we have set $\beta=0$ (from
(\ref{eq:Hcross3})) and used $\delta_{3e}=(2 \pi)1.5$~GHz, instead
of the $(2 \pi)10$~GHz we predict for our proposed parameters.  We
use $x_{14}=0.0054$ and $x_{24}=-0.0437$ and
$\Gamma_e=\Gamma_3^{(t)}=(2 \pi)159$~MHz.}
\end{figure}

The last important situation we consider is that of coupling to an
additional excited level $\ket{e}$, coupled off-resonantly to
states $\ket{1}$ and $\ket{2}$ via the two applied fields $a$ and
$b$, thus forming a ``double-$\Lambda$'' system \cite{fourLevel},
as diagrammed in Fig.~\ref{fig:crossTalk4level}(a).  As we will
see, the extra coupling gives rise to AC Stark shifts in much the
same way as we saw in Section~\ref{subsec:3LevelCross}.  In
addition, because $\ket{e}$ (unlike $\ket{2}$) is quickly
decaying, the coupling gives rise to some population loss even
when the AC Stark shift is compensated.

Dropping the RWA with respect to terms coupling to level $\ket{e}$
and defining $\delta_{3e} \equiv \omega_e-\omega_3$, the Hamiltonian
(using a frame $\delta_e=\Delta_{13}+\delta_{3e}$) is

\begin{align}
 {\cal{H}} = \frac{\hbar}{2} \left[\begin{matrix}
         0 & 0 & \Omega_{13}^* & \beta_1^* \Omega_{13}^* \\
     0 & - 2 \Delta_2 & \Omega_{23}^* & \beta_2^* \Omega_{23}^* \\
      \Omega_{13}& \Omega_{23} & -2 \Delta_{13} -  i \Gamma_3^{(\mathrm{t})} & 0 \\
      \beta_1 \Omega_{13} & \beta_2 \Omega_{23} & 0 &
     -2 (\Delta_{13}+ \delta_{3e}) -  i \Gamma_e
         \end{matrix}\right],
 \label{eq:Hcross4}
\end{align}

\noindent with $\beta_i \equiv x_{i4}/x_{i3}$.  We have assumed some
large open loss channel $\Gamma_e$.

In the case where only one of the couplings is present
($\beta_1=0$ or $\beta_2=0$), the effect is simple to calculate.
When $\beta_1=0$, one can consider the Schr\"odinger evolution
$\dot{c}_e$ from Eq.~(\ref{eq:Hcross4}) and adiabatically
eliminate $c_e$ to obtain

\begin{equation}
\label{eq:cE} c_e = -i \frac{\beta_2 \Omega_{23}c_2}{2 i
\delta_{3e}- \Gamma_e}
\end{equation}

\noindent  where we have assumed $\delta_{3e} \gg \Delta_{13}$.
Substituting this back into the equation for $\dot{c}_2$ yields:

\begin{equation}
\label{eq:c1evol} \dot{c}_2 = -\frac{i}{2} \Omega_{23} c_3 - i
\Delta_2 c_2- |\beta_2 \Omega_{23}|^2 \bigg(\frac{2 i
\delta_{3e}+\Gamma_e}{8 \delta_{3e}^2+2 \Gamma_e^2}\bigg)c_2
\end{equation}

\noindent revealing that the extra coupling gives rise to a Stark
shift and population decay of $\ket{2}$.  In the large detuning
limit ($\delta_{3e} \gg \Gamma_e$), the Stark shift is $|\beta_2
\Omega_{23}|^2/4\delta_{3e}$ and we have an effective loss rate from
$\ket{2}$, $\Gamma_2^{(e)}=|\beta_2 \Omega_{23}|^2 \Gamma_e/4
\delta_{3e}^2$. Analogous results occur when $\beta_2=0$, leading to
a Stark shift of the two-photon resonance of opposite sign. The
population loss rates will in turn contribute to the decoherence and
cause exponential loss from the EIT as discussed in
Section~\ref{sec:loss}.

When both are present, the shifts and loss rates are not simply
the sum of the two separate contribution, due to interferences
between them.  A case where summing the two contributions clearly
does not work is $\beta_1=\beta_2$ (equal dipole ratios).  In this
case (and {\it only} this case) the dark state $c_2/c_1 =
-\Omega_{13}/\Omega_{23}$ is also completely decoupled from
$\ket{4}$.  Thus no population loss or Stark shift is induced in
this case.  To obtain an expression in the general case, we follow
the following procedure.  We adiabatically eliminate $c_3$ and
$c_e$ from the Schr\"odinger equation from Eq.~(\ref{eq:Hcross4}),
obtain the $2 \times 2$ evolution matrix for $c_1$ and $c_2$, then
solve for the eignvalues of this matrix.  One of the values has a
large imaginary part (which reduced to
$\Omega^2/\Gamma_3^{(\mathrm{t})}$ in the limit
$\beta_1=\beta_2=0$) and corresponds to the absorbing state. The
other has a small imaginary part which vanishes when
$\beta_1=\beta_2=0$ and corresponds to the dark state.  We
investigated the imaginary part of this eigenvalue in the limit
that $\Omega^2/\Gamma_3^{(\mathrm{t})} \gg
\Delta_{13},\Delta_{23}$ and $\beta_i \Gamma_e/\delta_{3e} \ll 1$
and then minimized this expression with respect to the two-photon
detuning $\Delta_2$ (when $\Delta_{23}=0$) to obtain the Stark
shift. The loss rate and shift of the resonance obtained were:

\begin{eqnarray}
\label{eq:AClossE} R_L^{(e)} & = & \frac{|\Omega_{13}|^2
|\Omega_{23}|^2}{\Omega^4}
\frac{\Omega^2 \Gamma_e}{4 \delta_{3e}^2+\Gamma_e^2}|\beta_1-\beta_2|^2; \nonumber \\
\Delta_{AC}^{(e)} & = & -\frac{\delta_{3e}}
{4\delta_{3e}^2+\Gamma_e^2}(\beta_1-\beta_2)(\beta_1
|\Omega_{13}|^2+\beta_2 |\Omega_{23}|^2).
\end{eqnarray}

\noindent These expressions reduce to the simpler cases above
($\beta_1=0$ or $\beta_2=0$) and also disappear when
$\beta_1=\beta_2$.  Interestingly, the relative sign of the dipole
moments plays an important role.  For example if $\beta_1=-\beta_2$
the loss rate is actually twice what one would expect from the sum
of the individual couplings.  In this case, the dark state $\Psi_D$
for $\ket{3}$ is actually the absorbing state for $\ket{4}$.

In Fig.~\ref{fig:crossTalk4level}(b) we plot the population
remaining versus detuning $\Delta_{13}$ for three different field
intensities in the system we have been considering, but now
considering the coupling to the additional level $\ket{e}$. The
dots show numerical propagation of the density matrix equations
corresponding to Eq.~(\ref{eq:Hcross4}). Note that to isolate the
effect studied at present, we have ignored the cross-talk
considered in Section~\ref{subsec:3LevelCross} (by setting
$x_{12}=0$) and set the pure dephasing $\gamma_{12}=0$.  The solid
curves then show the analytic estimate based on the Stark shifts
and loss rates (\ref{eq:AClossE}). One sees excellent agreement.

It should be noted that this analysis should be able to account for
the effect of multiple excited levels ${e_j}$ by simply summing
their contributions $\sum_j R_L^{(e_j)}$ and $\sum_j
\Delta_{AC}^{(e_j)}$. Because of the large frequency differences
between each successive level, coherent interference between
contributions from different $\ket{e_j}$ will not occur.

\section{\label{sec:conclusion} Conclusion}

We have described in detail a proposal for demonstrating a quantum
optical effect, EIT, in a SQC.  In this context, EIT will manifest
itself as the suppression of photon-induced tunneling from stable
states $\ket{1},\ket{2}$ through some read-out state $\ket{3}$, due
to quantum mechanical interference for two paths of excitation. This
provides a method of unambiguously demonstrating phase coherence in
these systems.  We have provided a thorough and mostly analytic
treatment of EIT in the presence of complicating effects due to
decoherence and multiple levels in SQCs, which will be important in
guiding experimental implementation and observation of EIT and other
quantum intereference effects.

We analyzed in detail first the basic considerations of EIT such as
imperfect dark state preparation, and one- and two-photon detuning
and determined the expected experimental signatures. Under
appropriate conditions, we obtained an expression for the total
population as a function of time Eq.~(\ref{eq:pop}), which describes
a fast loss of the absorbing component, followed by a small
exponential loss of the system.  Over shorter times, EIT thus
provides a method to confirm the successful preparation (and
coherence) of the particular dark state defined by the microwave
fields applied.   For longer times, the observed loss rate will be
function of both the detuning from two-photon resonance as well as
decoherence effects. We also discussed the important issue of how
the measurement state $\ket{4}$ plays a role in the decay of the
read-out state $\ket{3}$ and saw how the biasing condition of these
levels and the SQUID measurement rate can have a large effect on the
parameters of the effective three-level system.

We then discussed in detail how decoherence due to dephasing of
the qubit coherence, incoherent population loss or exchange, and
tunneling of levels through the barrier effects the loss rate.
Measuring these loss rates can then be a powerful tool which
sensitively probes these various processes.  We obtained the
coefficients for the loss rates, which depend differently on the
field strengths, depending on the underlying decoherence
processes. For the case of primarily coherent resonant tunneling,
we found that that the EIT will exhibit a qualitatively different
double-peaked structure. Probing these effects with EIT can aid in
understanding and minimizing decoherence and give information
about the full multi-level structure of the SQC.  A potentially
interesting future investigation is to learn the signature from
coupling to other quantum degrees of freedom, such as the
microresonators postulated in \cite{microRes}.

Finally we have found that the microwave fields themselves can cause
additional loss rates and AC Stark shifts of the EIT resonance which
must be accounted for when one uses stronger field strengths.
Importantly, we found that, these effects can becomes more
pronounced with $\Omega^2$, meaning there will be some intermediate
field strength which balances these considerations with the
decoherence and detuning effects.  Also, we showed how these effects
could be mitigated by proper compensation of the Stark shifts.

This work was supported in part by the AFOSR grant
No.~F49620-01-1-0457 under the Department of Defense University
Research Initiative in Nanotechnology (DURINT). The work at
Lincoln Laboratory was sponsored by the AFOSR under Air Force
Contract No.~F19628-00-C-0002.  Z.D. acknowledges support from the
Office of Naval Research.

%\bibliography{qcmaster-murali}

\end{document}